\documentclass[aps,prd,showpacs,showkeys,preprintnumbers,notitlepage,superscriptaddress]{revtex4-1}
\usepackage{amsmath,amssymb,amsthm,bm,bbm,color,hyperref}
\usepackage{graphicx,subfigure,rotating}

\begin{document} 

\title{A model of random center vortex lines in continuous
2+1-dimensional space-time}

\author{Derar Altarawneh}
\email{derar@ttu.edu.jo}
\affiliation {Department of Applied Physics, Tafila Technical University, Tafila , 66110 , Jordan}
\author{Michael Engelhardt}
\email{engel@nmsu.edu}
\affiliation{Department of Physics, New Mexico State University, PO Box 30001, Las Cruces, NM 88003-8001, USA}
\author{Roman H\"ollwieser\footnote{Funded by an Erwin Schr\"odinger Fellowship of the Austrian Science Fund under Contract No. J3425-N27.}}
\email{hroman@kph.tuwien.ac.at}
\affiliation{Department of Physics, New Mexico State University, PO Box 30001, Las Cruces, NM 88003-8001, USA}
\affiliation{Institute of Atomic and Subatomic Physics, Vienna University of Technology, Operngasse 9, 1040 Vienna, Austria}

\date{\today}

\begin{abstract}
A picture of confinement in QCD based on a condensate of thick vortices
with fluxes in the center of the gauge group (center vortices) is studied.
Previous concrete model realizations of this picture utilized a
hypercubic space-time scaffolding, which, together with many advantages,
also has some disadvantages, e.g., in the treatment of vortex topological
charge. In the present work, we explore a center vortex model which does
not rely on such a scaffolding. Vortices are represented by closed random
lines in continuous 2+1-dimensional space-time. These random lines are
modeled as being piece-wise linear, and an ensemble is generated by
Monte Carlo methods. The physical space in which the vortex lines
are defined is a torus with periodic boundary conditions. Besides moving,
growing and shrinking of the vortex configurations, also reconnections
are allowed. Our ensemble therefore contains not a fixed, but a variable
number of closed vortex lines. This is expected to be important for
realizing the deconfining phase transition. We study both vortex
percolation and the potential $V(R)$ between quark and anti-quark as
a function of distance $R$ at different vortex densities, vortex
segment lengths, reconnection conditions and at different temperatures.
We find three deconfinement phase transitions, as a function of density,
as a function of vortex segment length, and as a function of temperature.
\end{abstract}



\maketitle

\tableofcontents

\newpage

\section{Introduction}
Quantum Chromodynamics (QCD) is the regnant theory of the strong interaction.
It is formulated in terms of quarks and gluons, which are the basic degrees
of freedom that make up hadronic matter. QCD is well understood in the
regime where we have a large momentum transfer (ultraviolet regime). In
this regime, the theory is weakly coupled and can thus be solved using
perturbative methods. On the other hand, at low energy, analytical
solutions are very hard to obtain due to the large coupling constant and
the highly nonlinear nature of the strong force. It happens especially
in this infrared regime that the QCD vacuum exhibits some extraordinary
features, among them the confinement of quarks into bound hadrons and
chiral symmetry breaking ($\chi$SB), the origin of mass in QCD.
A perspective to construct a cogent, comprehensive model of the strong
interaction vacuum in which, in particular, a connection between
topological properties and confinement can be drawn, appeared in the
framework of the magnetic (center) vortex picture~\cite{'tHooft:1977hy,Vinciarelli:1978kp,Yoneya:1978dt,Cornwall:1979hz,Mack:1978rq,Nielsen:1979xu}:
Chromo-magnetic flux lines compress the chromo-electric flux between
color electric sources into a flux tube (or a "string"), resulting in
a linearly rising potential and thus confinement.

In $D$-dimensional space-time, center vortices are (thickened)
($D-2$)-dimensional chromo-magnetic flux degrees of freedom. The center
vortex picture of the strong interaction vacuum assumes that these are
the relevant degrees of freedom in the infrared sector of the strong
interaction; the center vortices consequently are taken to be weakly
coupled and can thus be expected to behave as random lines (for $D = 3$)
or random surfaces (for $D = 4$). The magnetic flux carried by the
vortices is quantized in units which are singled out by the topology
of the gauge group, such that the flux is stable against small local
fluctuations of the gauge fields. In the vortex model of confinement,
the deconfinement transition results from a percolation transition
of these chromo-magnetic flux degrees of freedom. This theoretically
appealing picture has been buttressed by a multitude of numerical
calculations, both in lattice Yang-Mills theory and within a corresponding
infrared effective model, see {\it e.g.}~\cite{DelDebbio:1996mh,Langfeld:1997jx,DelDebbio:1997ke,Langfeld:1998cz,Kovacs:1998xm,Engelhardt:1999wr,Engelhardt:1999fd,Bertle:2002mm,Engelhardt:2003wm,Hollwieser:2014lxa,Altarawneh:2015bya,Hollwieser:2015qea},
or~\cite{Greensite:2003bk}, which summarizes the main features. 
Recent results~\cite{Greensite:2014gra} have also suggested that the center
vortex model of confinement is more consistent with lattice results than other
currently available models. Lattice studies further indicate that vortices
may also be responsible for topological charge~\cite{Bertle:2001xd,Engelhardt:2000wc,Engelhardt:2010ft,Hollwieser:2010mj,Hollwieser:2011uj,Schweigler:2012ae,Hollwieser:2012kb,Hollwieser:2014mxa,Hollwieser:2015koa}
and $\chi$SB~\cite{deForcrand:1999ms,Alexandrou:1999vx,Engelhardt:1999xw,Reinhardt:2000ck2,Engelhardt:2002qs,Leinweber:2006zq,Bornyakov:2007fz,Hollwieser:2008tq,Bowman:2010zr,Hollwieser:2013xja,Brambilla:2014jmp,Hollwieser:2014osa,Trewartha:2014ona,Trewartha:2015nna},
and thus unify all non-perturbative phenomena engendered by the structure
of the strong interaction vacuum in a common framework. 

A concrete implementation of the vortex picture, using a hypercubic lattice
scaffolding to support the random vortex lines or surfaces, has been studied
extensively by one of the
authors~\cite{Engelhardt:1999wr,Engelhardt:2000wc,Engelhardt:2002qs,Engelhardt:2003wm,Quandt:2004gy,Engelhardt:2004qq,Engelhardt:2005qu,Engelhardt:2006ep,Engelhardt:2010ft}.
The hypercubic formulation has a number of advantages, among them, simple
Monte Carlo updates which naturally include surfaces fusing and
disassociating, and a straightforward bookkeeping of vortex location,
permitting, {\it e.g.}, simple evaluation of Wilson loops and counting of
vortex surface intersections. On the other hand, however, this formulation
has revealed weaknesses as far as the calculation of topological charge is
concerned. Vortex world-surface configurations in this formulation, being
restricted to only six discrete space-time directions in which they can
extend, exhibit ambiguities in the definition of topological charge
which would not appear in ensembles of arbitrary two-dimensional surfaces
in continuous four-dimensional space-time. 

In view of this, we explore in the present work an alternative formulation,
which avoids the shortcomings of the hypercubic construction, concentrating
in a first step on a model of random flux lines in $D=2+1$ space-time
dimensions, representing vortices of the $SU(2)$ gauge group (i.e., there are
no branchings of the vortex lines \cite{Engelhardt:1999wr,Engelhardt:2003wm}).
The lines are composed of straight segments connecting
nodes randomly distributed in three-dimensional space. Allowance is made for
nodes moving as well as being added or deleted from the configurations
during Monte Carlo updates. Furthermore, Monte Carlo updates disconnecting
and fusing vortex lines, {\it i.e.}, reconnection updates are implemented.
Given that the deconfining phase transition is a percolation transition,
such processes play a crucial role in the vortex picture. The model is
formulated in a toroidal finite volume, with periodic boundary conditions,
which allows for a study of finite temperatures (via changes in the temporal
extent of the volume). The resulting vortex ensemble is used, in
particular, to evaluate the string tension and its behavior as a function
of temperature, with a view towards detecting the high-temperature
deconfining phase transition.

The above scheme of modeling random lines (and higher-dimensional manifolds)
is reminiscent of models employed in the study of quantum gravity~\cite{Baillie:1989mv,Catterall:1995:SDT,Jain:1992bs,Ambjorn:1993vz,Thorleifsson:1995ki,Ambjorn:1995dj}. While the present work focuses only on the lowest-dimensional
case, the inclusion of a variable number of (vortex) clusters in the ensemble
is a feature that is not generally contemplated in quantum gravity
applications. Here, it is crucial in order to include the physics of
the deconfinement transition. Also the use of a torus with periodic
boundary conditions, on which the vortices are defined, in order to
treat finite temperature, constitutes a significant complication.

This paper is organized as follows: The modeling details will be given
in section~\ref{sec:model}, observables will be introduced in
section~\ref{sec:measo}, and results will be presented and
discussed in section~\ref{sec:res}. Section~\ref{sec:concl} provides a
summary of the main results and a short outlook.

\section{The Model}\label{sec:model}
\begin{figure}[h]
\centering
\includegraphics[width=.8\linewidth]{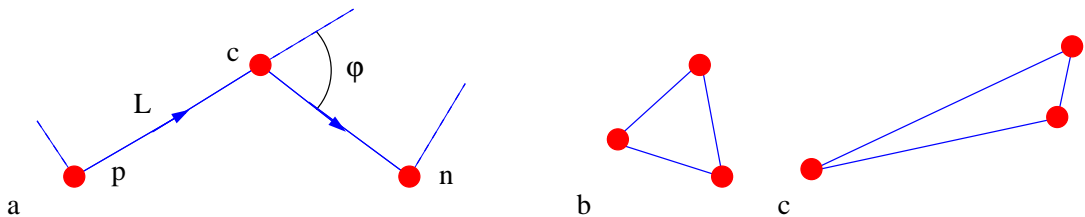}
\caption{a) The action $S=\alpha L + \gamma \varphi^2$ of the current node (c) is given by the vortex length $L$ to the previous node (p) and the angle $\varphi$ between the vortex lines to the previous (p) and the next node (n). Three node clusters of shape (b) are more likely accepted by the Metropolis algorithm than of shape (c).}\label{fig:act}
\end{figure}

In the model, vortices are represented by closed random lines in
2+1-dimensional (Euclidean) space-time. The physical space in which the
vortex lines are defined is a torus $L_S^2\times L_T$ with "spatial"
extent $L_S$, "temporal" extent $L_T$ and periodic boundary conditions
in all directions. The random lines are modeled as being
piece-wise linear between "nodes" with vortex segment length $L$ restricted
to a certain range $L_{min}<L<L_{max}$. This range sets a scale of
the model; for practical reasons we choose a scale of $L\approx1$, {\it i.e.}
$L_{min}=0.3$ and $L_{max}=1.7$ in appropriate units. Within this paper
we use volumes with $L_S=16$, where finite size effects are under control, and
varying time lengths $L_T$. Variations of the vortex segment length range
away from the aforementioned range will also be examined. An ensemble is
generated by Monte Carlo methods, starting with a random initial
configuration. A Metropolis algorithm is applied to add, move and
delete nodes using the action
\begin{equation}
	S= \alpha L + \gamma \varphi^2 \ ,\label{eq:act}
\end{equation}
with action parameters $\alpha$ and $\gamma$ for the vortex segment length
$L$ and the vortex angle $\varphi$ at nodes, respectively. At a given
(current) node the vortex segment length $L$ is defined to be the distance
to the previous node and the vortex angle is the angle
between the oriented vectors of the vortex lines connecting the previous,
current and next nodes, see Fig.~\ref{fig:act}. This type of action,
penalizing both vortex length and curvature, is analogous to the action
used in previous hypercubic lattice models
\cite{Engelhardt:1999wr,Engelhardt:2003wm}. Furthermore, when two
vortices approach each other, they can reconnect or separate at a bottleneck,
as described in detail further below. The ensemble therefore
will contain not a fixed, but a variable number of closed vortex lines or
"vortex clusters". This is expected to be important for realizing the
deconfining phase transition. Moreover, new (three-node) clusters are
permitted to "pop out" of the vacuum at random positions, again governed
by the above action; hence a small equilateral triangle is
more probable than a long acute triangle, see Fig.~\ref{fig:act}. The
new cluster then evolves further in subsequent updates, along with all
other clusters. Also the reverse process, annihilation of a three-node
cluster, is possible; it occurs when a node is deleted from a three-node
cluster, leading to a two-node cluster. Such a cluster is equivalent
to the absence of any flux, and is therefore deleted completely.

In the following, we discuss the individual updates and parameters of
the model in more detail.

\subsection{Move, Add \& Delete}\label{sec:upd}
\begin{figure}[h]
\centering
\includegraphics[width=.9\linewidth]{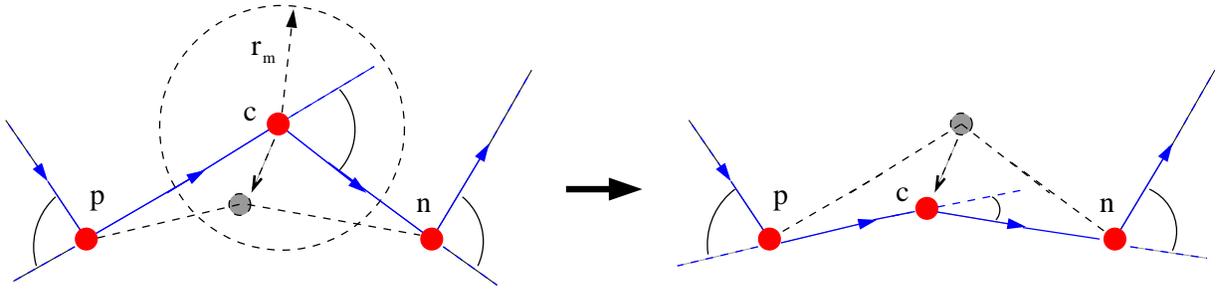}
\caption{The movement of the current node (c) within a certain range $r_m$ is
affecting the connected vortex lines and the angles at the current (c), the
previous (p) and the next node (n).}\label{fig:mov}
\end{figure}

Move, add and delete updates are applied to the vortex nodes via the
Metropolis algorithm, {\it i.e.}, the difference of the action of the affected
nodes before ($S_i$) and after ($S_f$) the update determines the probability
$P=\mbox{min} (1,\exp{(S_i-S_f)} )$ of the update being accepted. The move
update attempts to move the current node by a random vector of maximal
length $r_m=4L_{min}$; it affects the action of three nodes, the current node
itself and its neighbors, see Fig.~\ref{fig:mov}.

\begin{figure}[h]
\centering
\includegraphics[width=.9\linewidth]{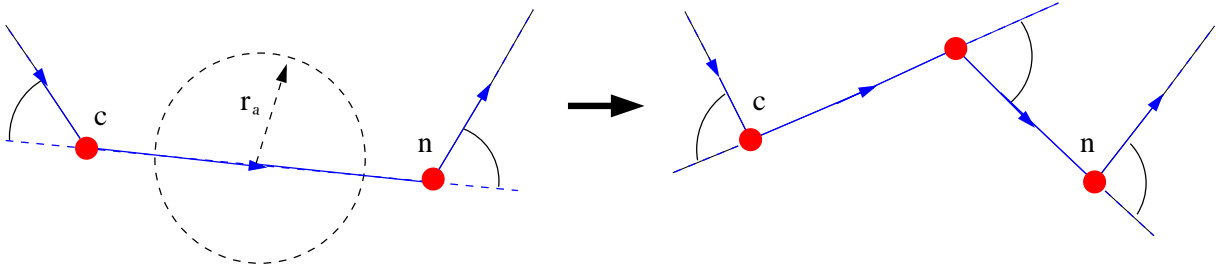}
\caption{The add update adds a node after the current node (c) within a
range $r_a$ around the midpoint of the vortex line to the next node (n).
Two vortex angles before and three after the update are affected, and one
vortex line is split into two.}\label{fig:add}
\end{figure}

\begin{figure}[h]
\centering
\includegraphics[width=.9\linewidth]{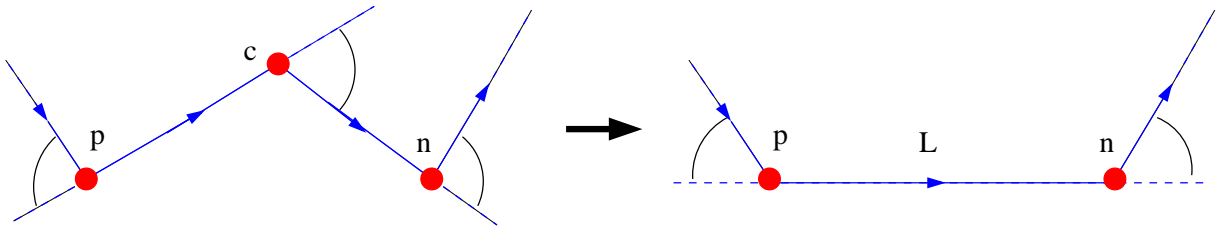}
\caption{The delete update deletes the current node (c), joining the connected
vortex lines into one between the previous (p) and the next node (n). It affects
three and two vortex angles before and after the update, respectively. The
new vortex length $L$ has to lie in the range $L_{min}<L<L_{max}$, like for
all other updates.}\label{fig:del}
\end{figure}

The add update attempts to add a node at a random position within a radius
$r_a=3L_{min}$ around the midpoint between the current and the next node, see
Fig.~\ref{fig:add}. The action $S_i$ before the update is given by the sum of
the action at the current and the next node, while the action $S_f$ after the
update is the sum of the action at the current, the new and the next node.
Conversely, deleting the current node affects three nodes, {\it i.e.}
the previous, the current and the next node before the update and only two nodes
(previous and next) after the update, see Fig.~\ref{fig:del}. Therefore the
probability $P$ for the add update is in general much smaller than for the
delete update; the vortex structure tends to vanish quickly if both updates
are tried equally often. As detailed below, add updates were attempted at
a significantly larger rate than delete updates.

In general, move, add and delete updates can come into conflict with the
restricted range of allowed vortex segment lengths $L$. In early
implementations, the move update was applied to every node of the
configuration, and if the resulting vortex segment lengths conflicted
with the allowed range, the corresponding nodes were deleted, or
auxiliary ones at midpoints were added, respectively. In principle, in
this scheme, the configuration can be stabilized for a set of fine-tuned
parameters, but these parameters lead to very dense vortex configurations
deep in the confinement phase. Hence, in order to explore the whole
phase space of the model, an additional density parameter $d$ is
introduced, restricting the number of nodes in a certain volume. The add update
is rejected if the number of nodes within a $3\times3\times3$ volume around the
new node exceeds the density parameter $d$. Also new clusters popping up
are subjected to this density cutoff, {\it i.e.}, the number of nodes in
a $3\times3\times3$ volume has to be less than $d-3$ for a three-node
cluster to pop up there. Further, all updates resulting in vortex segment
lengths $L$ out of the range $L_{min}<L<L_{max}$ are also rejected.
The update strategy is randomized to move a node in two out
of three cases ($66\%$), and apply the add update about five times more often
than the delete update ($28\%$ vs. $6\%$). Maximal movement and add "radii"
$r_m$ and $r_a$ are set to four resp. three times $L_{min}$. The different
parameters and restrictions in the model may seem artificial at first sight,
but they are optimized in order to guarantee a balance between action
and entropy of the system. The influence of the individual parameters
on the model and their "physical" effect to favor either action or entropy
will be discussed in Sec.~\ref{sec:res}. An overview of all parameters
and the Monte Carlo sweep will be given in sections~\ref{sec:par}
and~\ref{sec:mcs} respectively, but before this, a detailed discussion
of the reconnection update, which is applied after every move and add
update, is in order.

\subsection{Reconnections}\label{sec:rec}
\begin{figure}[h]
\centering
\includegraphics[width=.9\linewidth]{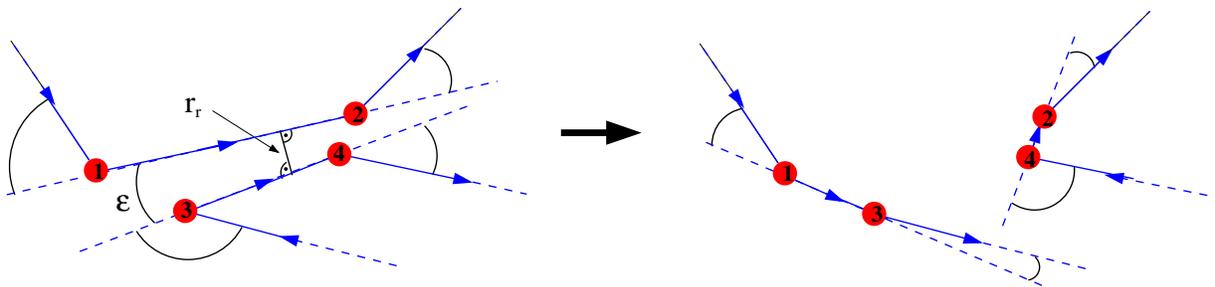}
\caption{The reconnection update deletes the vortex lines between
nodes 1-2 and 3-4 and reconnects nodes 1-3 and 4-2. The plot shows the affected
vortex angles, the reconnection angle $\epsilon$ and distance $r_r$. The
plotted vortex lines/nodes might belong to the same or different clusters.
Note that the orientation of some vortex lines was reversed in this
example, cf.~main text.}
\label{fig:rec}
\end{figure}

If the current node is not deleted, all nodes in a $3\times3\times3$
volume around the current node are considered for reconnections. The
reconnection update causes the cancellation of two close, parallel vortex
lines and reconnection of the involved nodes with new vortex lines.
Physically, this implements the fact that two vortex lines lying
on top of one another is equivalent to no physical flux being present;
since the actual vortices being represented are considered to possess
a certain thickness, the cancellation can be considered to occur as soon
as the vortices significantly overlap, i.e., are sufficiently close and
parallel. The terms "close" and "parallel" call for two more parameters in the
model, the reconnection length $r_r$ and the reconnection angle $\epsilon$. The
shortest distance and the angle between two vortex lines must be smaller than
$r_r$ and $\epsilon$ respectively, in order to reconnect the four involved nodes
with new vortex lines. An illustrative example of the reconnection update is
shown in Fig.~\ref{fig:rec}. The lengths of the new vortex lines, as always,
have to be smaller than $L_{max}$; however, the constraint of minimal
distance $L_{min}$ is not enforced in reconnection updates in order to
allow for reconnections of almost congruent vortex lines. If all
conditions for the reconnection are fulfilled, the update is subjected to the
Metropolis algorithm, considering the action of the four nodes involved. The
reconnection update allows separation and merging of vortex clusters. It should
be noted that the data structure of the vortex nodes imposes an orientation
on the vortex clusters (previous, next, etc.). If two merging clusters have
opposite orientation, the orientation of one of the clusters is reversed,
cf.~Fig.~\ref{fig:rec}. This orientation is a technical issue only and has
no physical meaning in this model.
 
\subsection{Parameters}\label{sec:par}
This section summarizes all parameters used in the model and the optimized
values used for the simulations on $16^2\times L_T$ volumes. It should be
noted that the search for viable parameter sets and update conditions
constituted the most demanding part of the simulation effort in this work.
This includes tuning for useful acceptance rates for move, add, delete,
and reconnection updates as well as new clusters popping up out of the
vacuum. In addition, a substantive competition between action and entropy
in the ensemble must be maintained to obtain physically interesting
behavior.

\begin{itemize}
\item Vortex length action parameter $\alpha=0.11$\\
	In the action $S=\alpha L+\gamma\psi^2$, the vortex segment length $L$ at a node is defined as the distance to the previous node.
\item Vortex angle action parameter $\gamma=0.33$\\
	The vortex angle $\varphi$ is defined as the angle between the oriented vectors of the vortex lines connecting the previous, current and next nodes, see Fig.~\ref{fig:act}.
\item Maximal vortex segment length $L_{max}=1.7$
\item Minimal vortex segment length $L_{min}=0.3$\\
	This parameter acts as a minimal length scale in the model, also determining
	\begin{itemize}
	\item the maximal radius of the move update $r_m=4L_{min}$, see Fig.~\ref{fig:mov}.
	\item the maximal radius of the add update $r_a=3L_{min}$, see Fig.~\ref{fig:add}.
	\item the reconnection length $r_r=L_{min}$, see Fig.~\ref{fig:rec}.
	\end{itemize}
\item Recombination angle $\epsilon=5^\circ$\\
	$\epsilon$ is the maximal angle between recombining vortex lines, see Fig.~\ref{fig:rec}.
\item Vortex density cutoff $d=8$\\
        Maximal number of nodes in a $3\times3\times3$ volume.
\end{itemize}

\subsection{Monte Carlo Algorithm}\label{sec:mcs}
For a simulation, the following algorithm was executed a total number
of $n_r=n_w+n_m*n_s$ times, where $n_w=10^4$ is the number of equilibration
iterations, $n_m=2\ldots5\times10^5$ is the number of
measurements and $n_s=10$ the number of sweeps between the measurements.
\begin{itemize}
	\item The Metropolis algorithm for one 3-node cluster pop-up is called
		before the node updates; therefore, the new nodes will also be
                updated before any measurement.
	\item Monte Carlo sweep over all nodes in the configuration:
	\begin{itemize}
		\item Metropolis move, add or delete updates are applied to the nodes with rates $66\%, 28\%$ and $6\%$.
		\item If a node is not deleted, possible reconnections are
			considered.
	\end{itemize}
	\item After the $n_w=10^4$ equilibration iterations are complete,
measurements are performed separated by $n_s=10$ Monte Carlo sweeps.
\end{itemize}
The next section introduces the observables measured in the model.

\section{Observables}\label{sec:measo}
The most directly accessible observables in the model are ones associated
with the action used to generate the ensemble, e.g., the total action itself
and the actual vortex density. These were used to analyze the equilibration
phase of the simulations. A number of $n_w=10^3$ equilibration sweeps was seen
to be generally sufficient for the model to thermalize; however, for the
following simulations, $n_w=10^4$ thermalization steps were used. After that,
the average action per node, the actual vortex density, the average vortex
segment length and angle, and Wilson loops were measured, and a vortex cluster
analysis was performed, every $n_s=10$ Monte Carlo sweeps. Wilson loop
and vortex cluster measurements will be detailed below. The vortex density
is a nontrivial observable since the cutoff parameter $d$ is applied only
for the add update; the vortex node density can locally exceed this
cutoff since the vortex nodes can move without density restrictions.
The actual vortex density is then given by the node density times the
average vortex segment length.

\subsection{Vortex Cluster Analysis}
The vortex cluster analysis comprises counting the number of closed vortex
clusters, the number of vortex nodes/line segments for each cluster, the
cluster size or maximal extent of each cluster and the number of clusters
winding around the time dimension. The distribution of vortex flux into
clusters of different sizes will be visualized in cluster size histograms
binning vortex nodes into 20 bins corresponding to the sizes of the clusters
to which they belong, where cluster size is normalized using the maximal
possible cluster size $s_m$. Taking into account the periodic boundary
conditions, the maximal possible cluster size is determined by
$s_m^2=L_S^2/2+L_T^2/4$. In the following analysis, the expression
"maximal cluster fraction" will refer to the fraction of vortex
nodes/line segments which reside within a cluster of size $s_m$, {\it i.e.}
the magnitude of the bin of maximal possible cluster size in the
aforementioned histograms. This quantifies to what extent vortices percolate.
On the other hand, vortex clusters winding around the time dimension are
important in particular in the deconfined phase; in the percolation
transition separating the confining and the deconfined phase, the large
percolating clusters of the confining phase decay into many such winding
clusters \cite{Engelhardt:1999wr}. The latter are instrumental in
maintaining a spatial string tension in the deconfined phase while
the physical string tension extracted from temporal Wilson loops
vanishes \cite{Engelhardt:1999wr}. Monitoring in particular vortices
winding around the time dimension therefore provides an additional
diagnostic for the deconfining transition. Such vortices can be produced
during reconnection updates, either in pairs or even singly if the
temporal extent of the torus is sufficiently small ($L_T<3L_{max}$).

\subsection{Wilson Loops}
The Wilson loop $W(R,T)$ along a closed rectangular path in space and time
of dimensions $R\times T$ is the observable most frequently used to discuss
confinement in lattice gauge theory. It can be interpreted in terms of the
creation of a static quark--anti-quark pair with a certain spatial
separation $R$, its evolution for a time $T$, and its subsequent
annihilation. The effective action associated with this process yields
the potential energy contained in the static quark--anti-quark pair.
Center vortices have a characteristic effect on Wilson loops; each center
vortex linked with a Wilson loop (or, equivalently, piercing any area
spanned by the loop) contributes a multiplicative factor to the loop
corresponding to a center element of the gauge group. This can, indeed,
be viewed as the defining property of a center vortex; it specifies the
flux carried by the vortex, which is measured by a Wilson loop encircling
it. In the case of the $SU(2)$ gauge group considered here, the only
non-trivial center element is $-1$; this is the factor by which any
Wilson loop linked to a vortex is multiplied.

To evaluate Wilson loops in the present model, it is sufficient to examine
all vortex line segments in a configuration, determining whether each line
segment pierces the planar area spanned by the Wilson loop in question,
and supplying a factor $-1$ to the Wilson loop for each piercing (if there
are no piercings, $W(R,T)=1$). Using the fact that larger Wilson loops are
simply given by products of smaller Wilson loops with which the larger loop
can be tiled, one can organize the calculation of a large number of Wilson
loops on a given configuration efficiently. The expectation value of the
time-like Wilson loops $\langle W(R,T)\rangle$ yields the
quark--anti-quark potential,
\begin{equation}
V(R) = -\lim_{T\rightarrow\infty} \;\ln\langle W(R,T)\rangle /T \ . 
\end{equation}
To extract the string tension $\sigma$ of the system, an ansatz $V(R)=\sigma
R+C/R+V_0$ is fitted to the potentials. The spatial string tension $\sigma_s$
is obtained from spatial Wilson loops using Creutz ratios,
\begin{equation}
\chi(R)=-\ln\left(\frac{\langle W(R,R)\rangle
\langle W(R+1,R+1)\rangle }{\langle W(R+1,R)\rangle \langle W(R,R+1)\rangle }
\right) \overset{R\rightarrow\infty}{\longrightarrow}\sigma_s \ .
\end{equation}
The spatial string tension is expected to be correlated with the number of
vortex clusters winding in time direction, since these vortices will pierce
the spatial Wilson loops.

\section{Results \& Discussion}\label{sec:res}

\subsection{Finite temperature phase transition from varying temporal
extent $L_T$}
\begin{figure}[h]
	\centering
	a)\includegraphics[width=\linewidth]{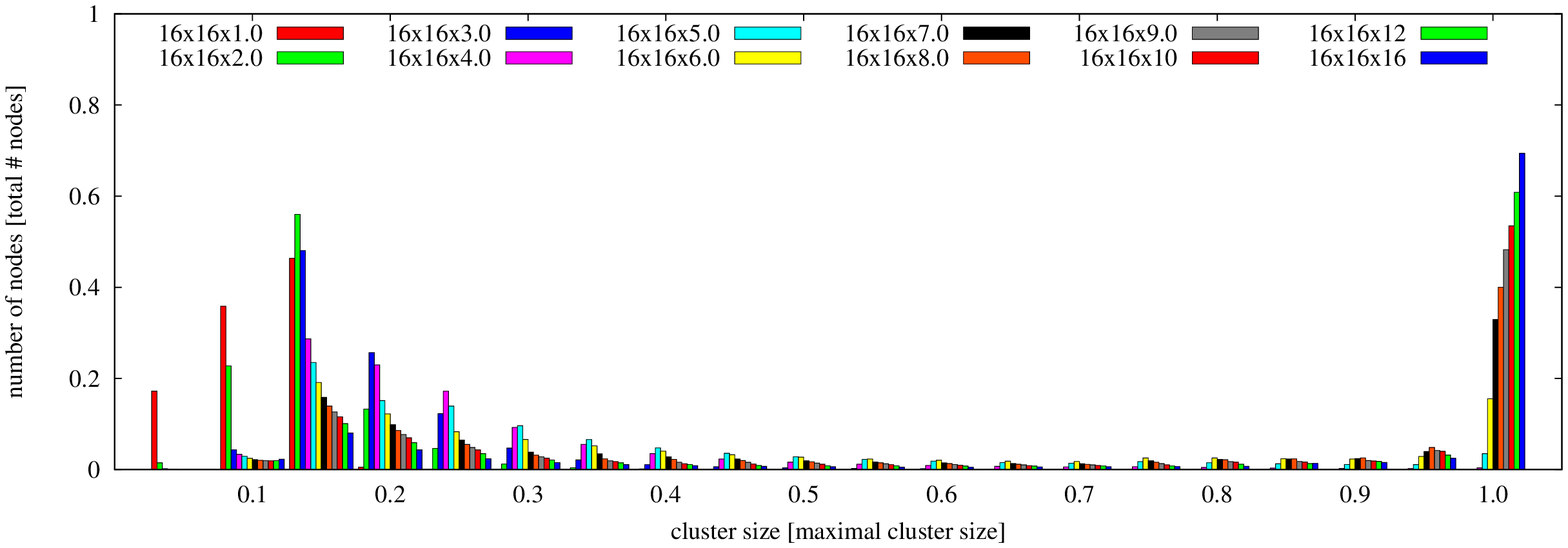}
	\vspace{3mm}\\
	b)\includegraphics[width=.48\linewidth]{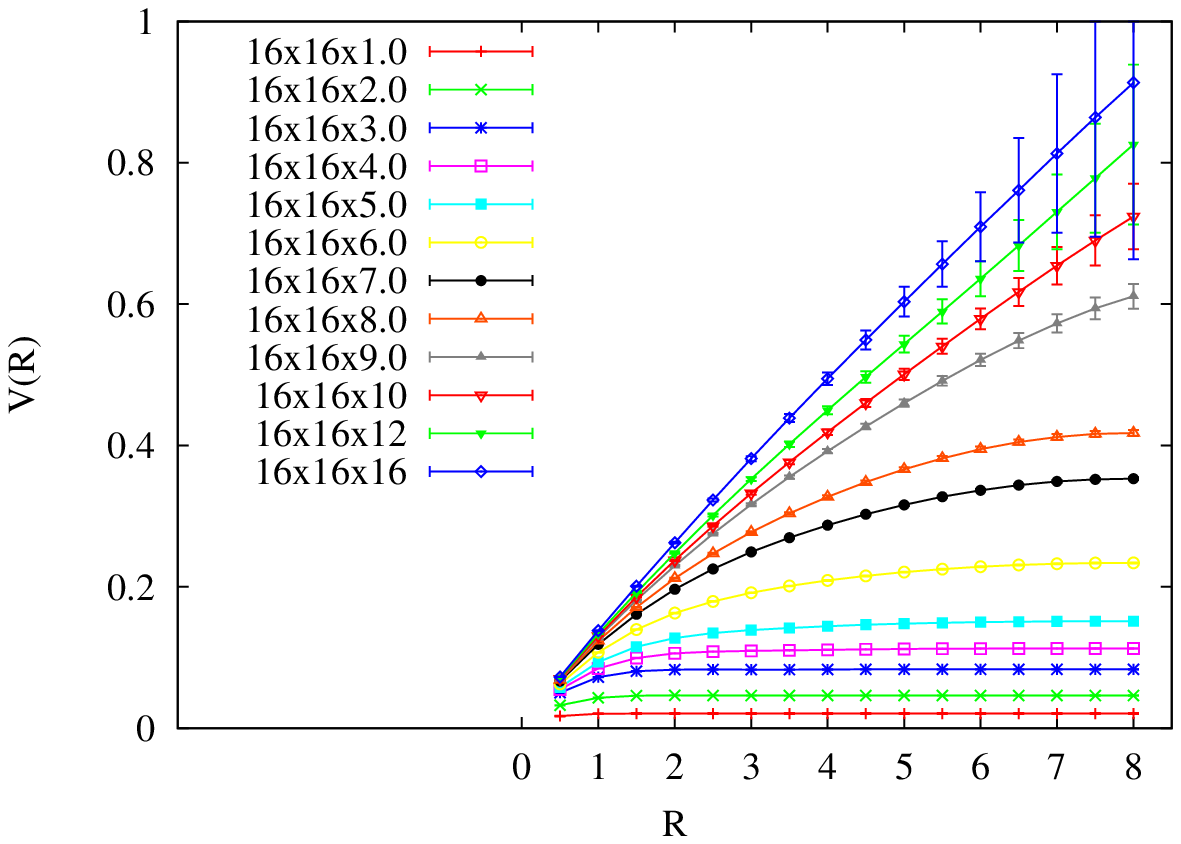}
	c)\includegraphics[width=.48\linewidth]{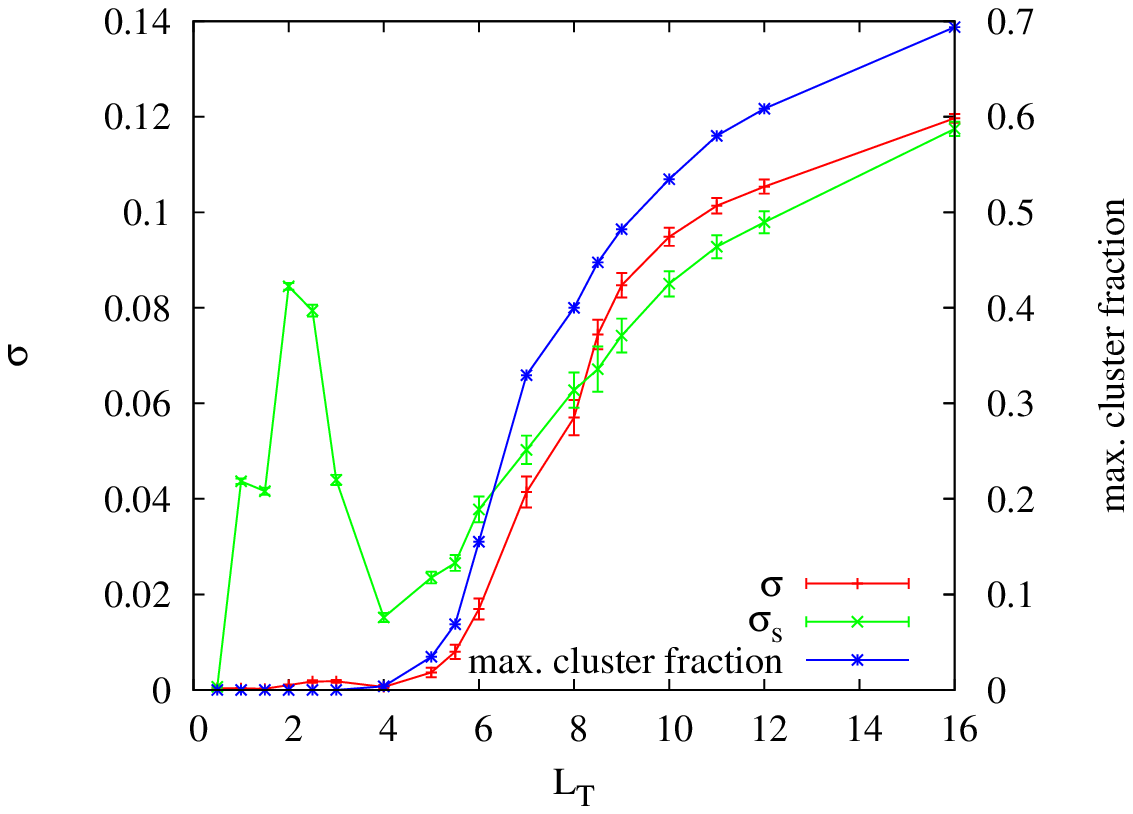}
	\caption{a) Cluster size histogram, b) quark--anti-quark potentials and c)
	maximal cluster fraction, temporal and spatial string tensions $\sigma$ and $\sigma_s$ for $16^2\times L_T$ volumes, density cutoff $d=4$, vortex lengths $L_{max}=1.7, L_{min}=0.3$.}
\label{fig:d4}
\end{figure}

\begin{figure}[h]
	\centering
	a)\includegraphics[width=.48\linewidth]{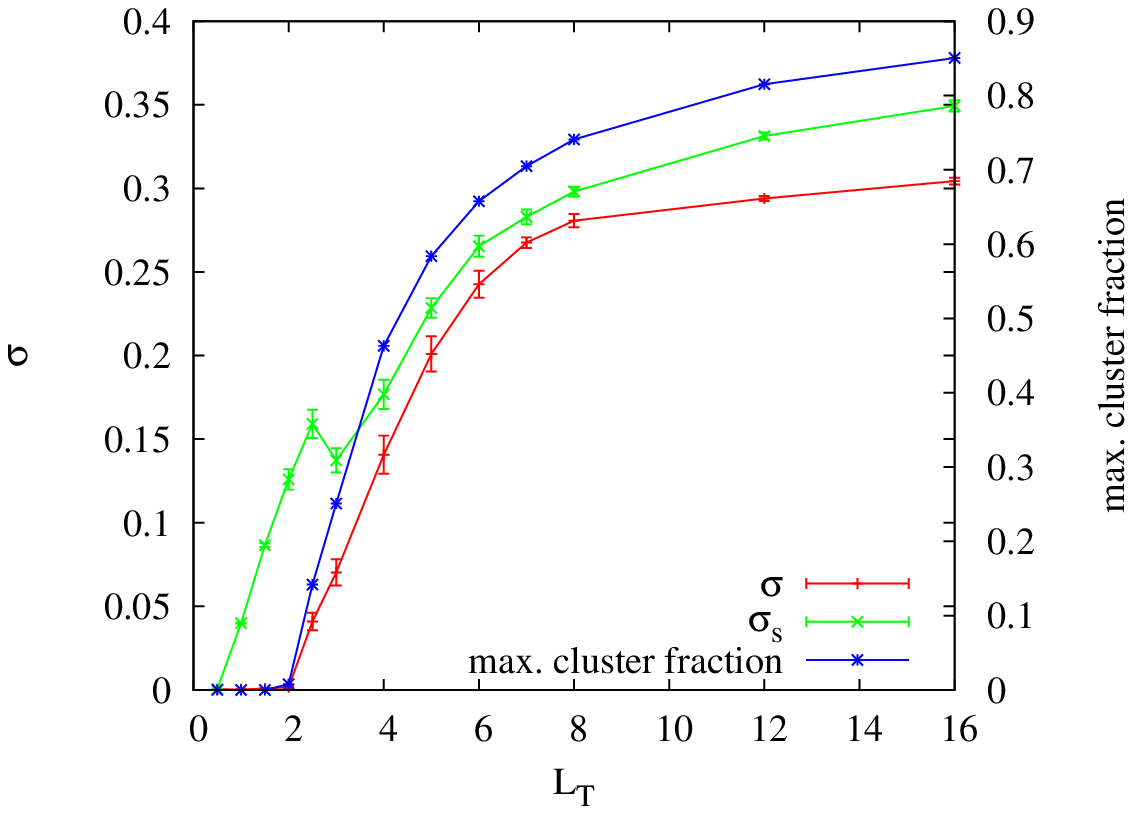}
	b)\includegraphics[width=.48\linewidth]{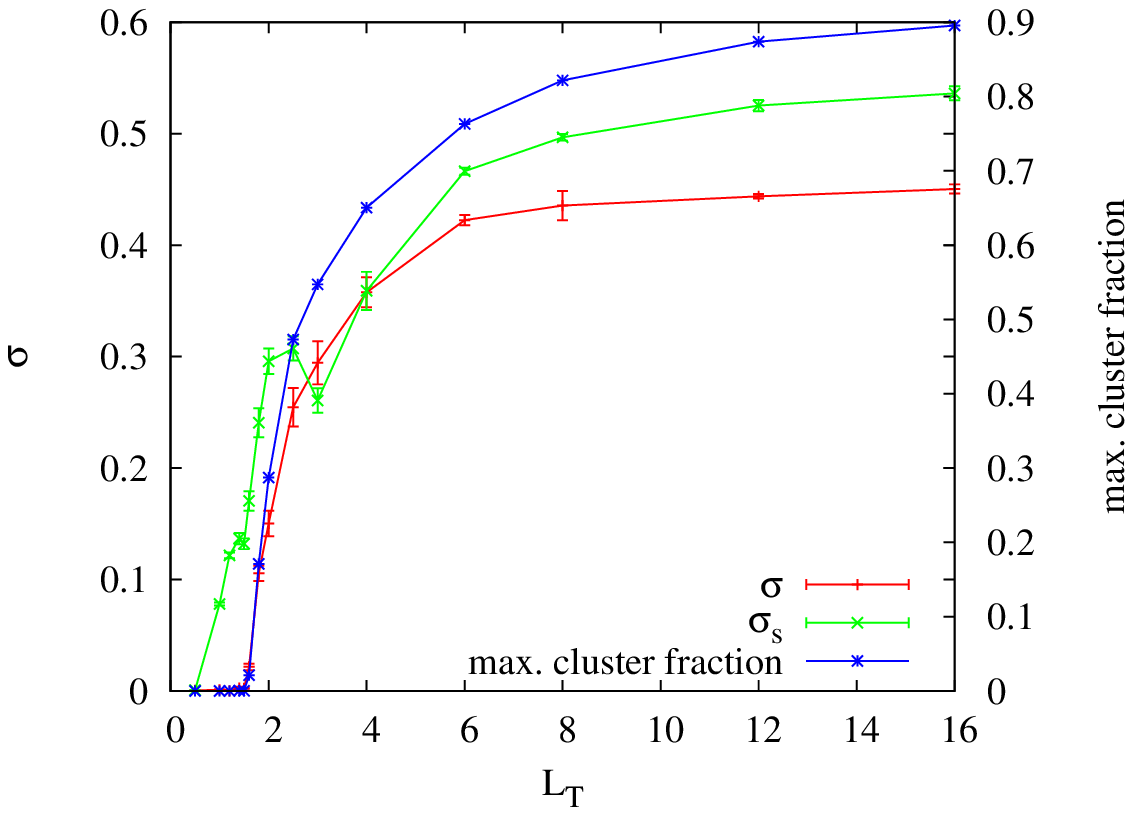}
	c)\includegraphics[width=.48\linewidth]{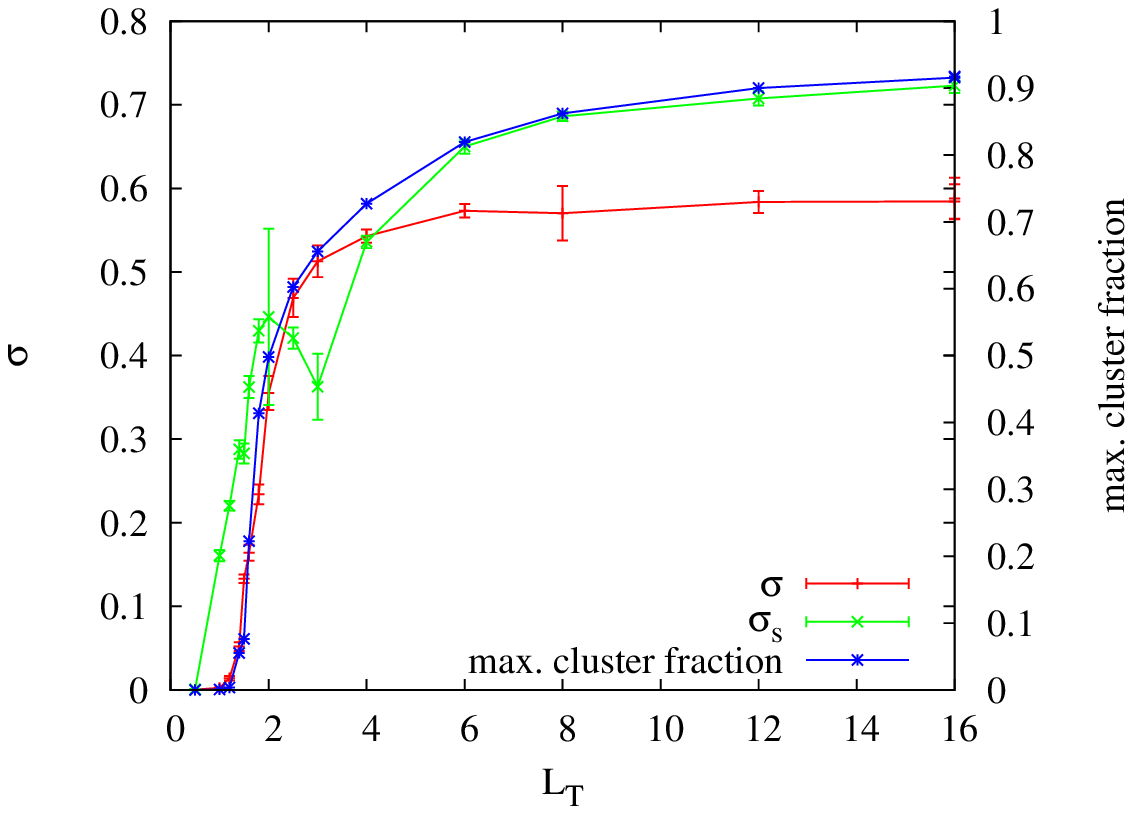}
	d)\includegraphics[width=.48\linewidth]{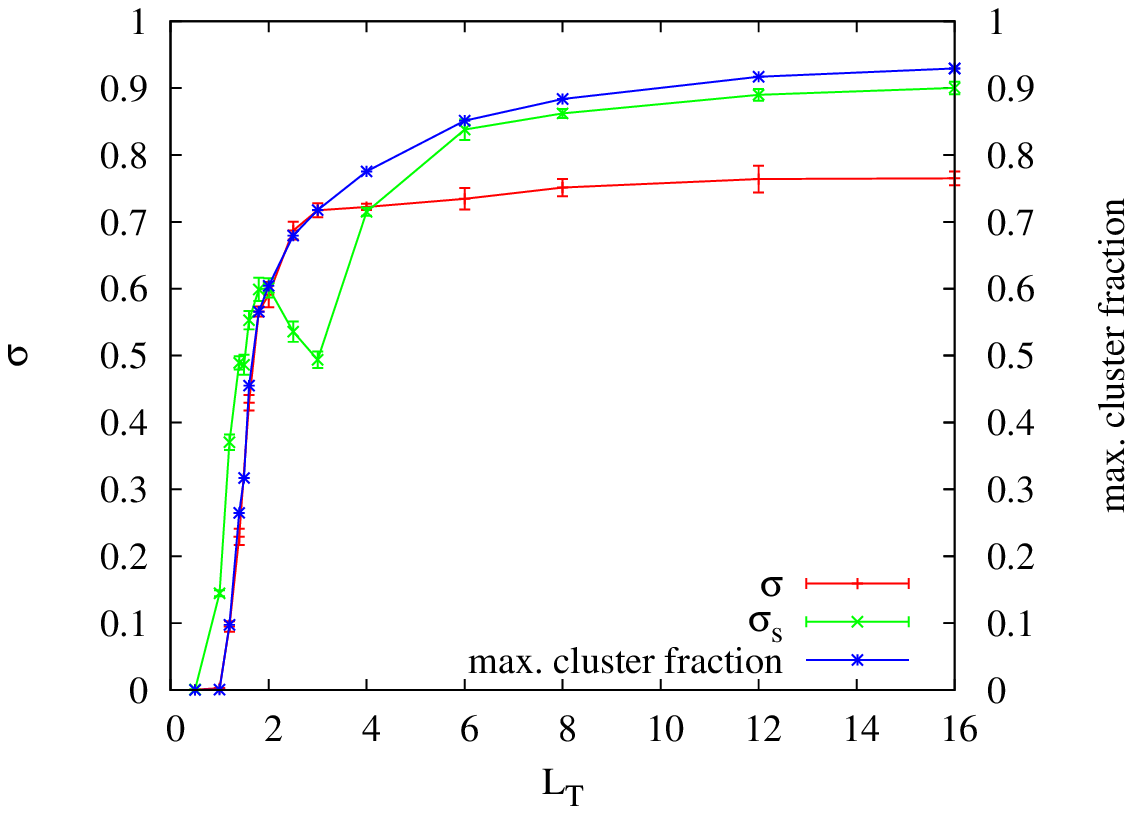}
	\caption{Maximal cluster fraction, temporal and spatial string tensions $\sigma$
	and $\sigma_s$ for $16^2\times L_T$ volumes, density cutoff a) $d=6$, b)
	$d=8$, c) $d=10$, d) $d=12$, vortex lengths $L_{max}=1.7, L_{min}=0.3$.}
\label{fig:sigT}
\end{figure}

\begin{figure}[h]
	\centering
	a)\includegraphics[width=.9\linewidth]{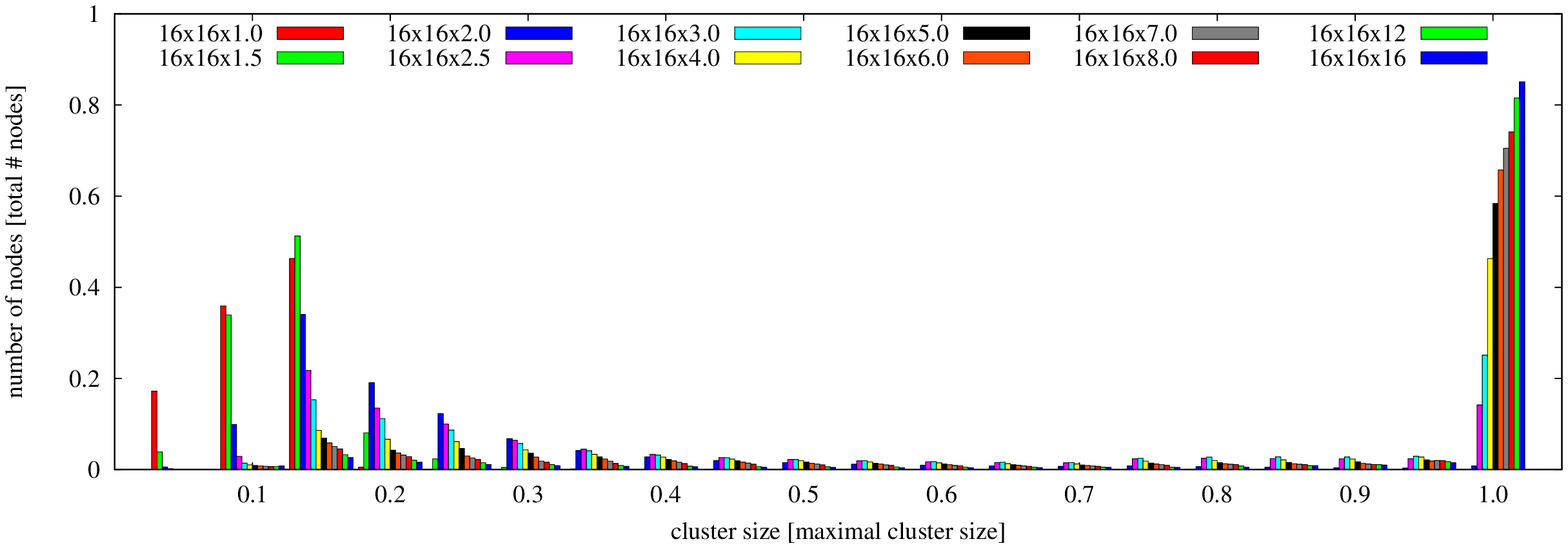}
	b)\includegraphics[width=.9\linewidth]{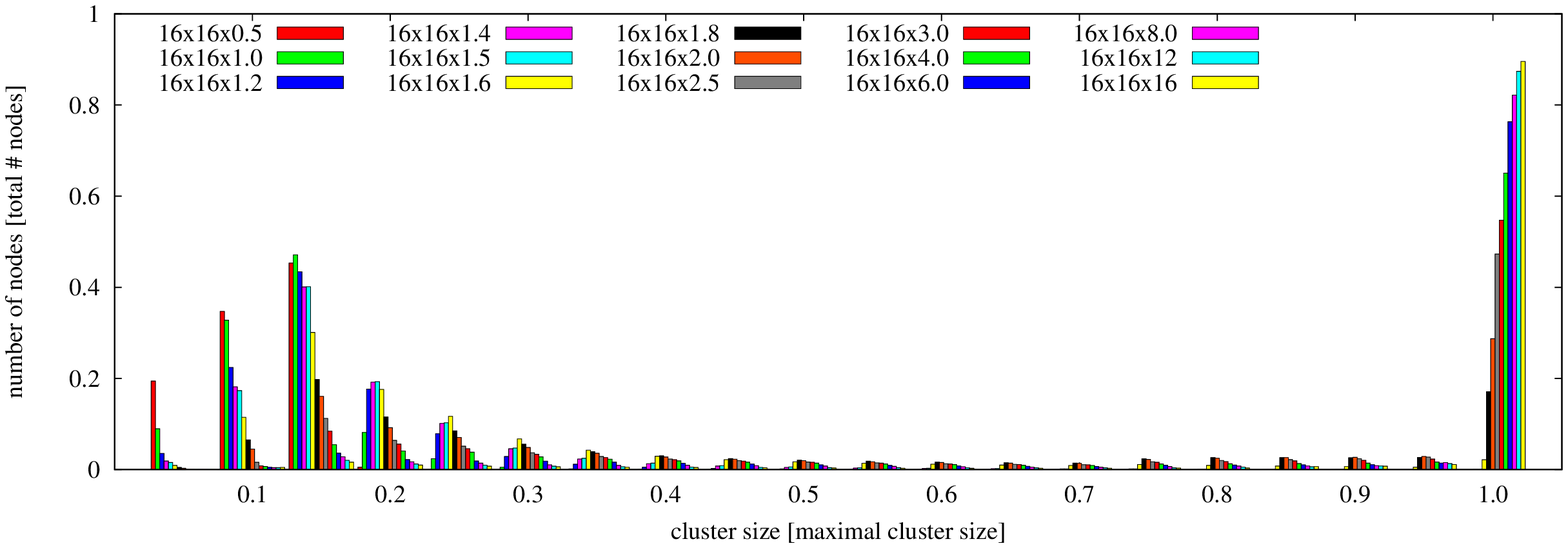}
	c)\includegraphics[width=.9\linewidth]{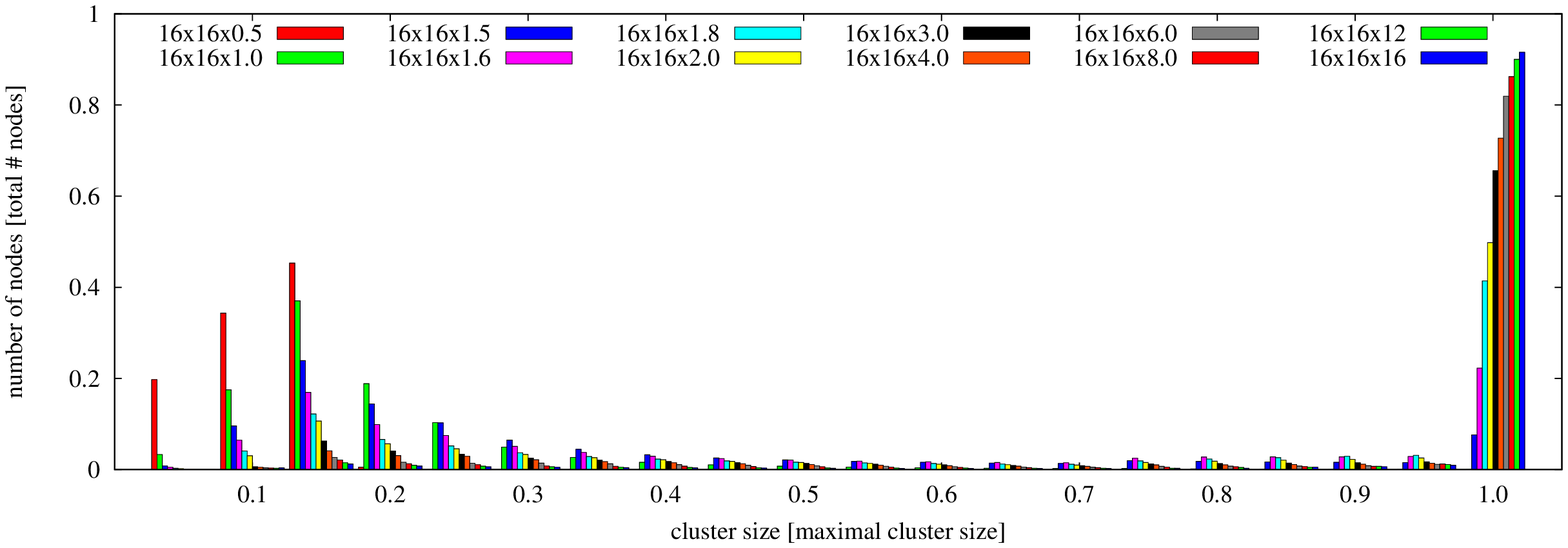}
	d)\includegraphics[width=.9\linewidth]{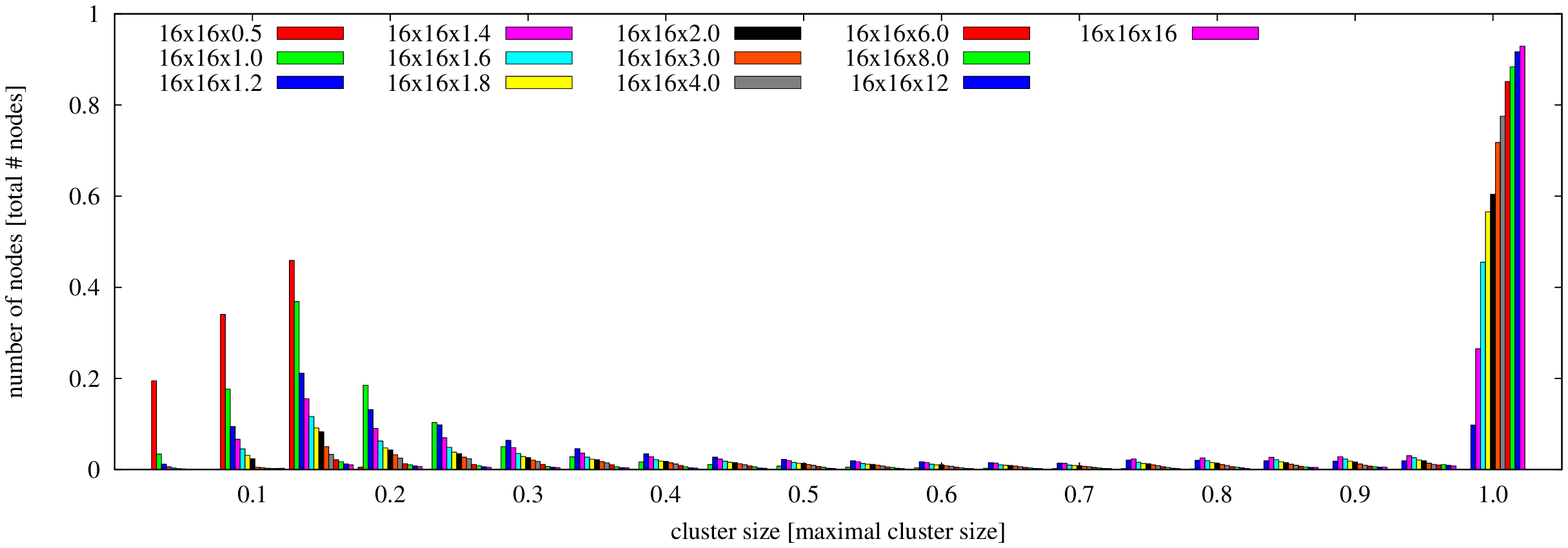}
	\caption{Cluster size histograms for $16^2\times L_T$ volumes, density cutoff a) $d=6$, b)
	$d=8$, c) $d=10$, d) $d=12$, vortex lengths $L_{max}=1.7, L_{min}=0.3$.}
\label{fig:hsT}
\end{figure}

In this section, we study center vortex ensembles at different temperatures.
The following results were obtained on volumes $16^2\times L_T$
for a range of inverse temperatures $L_T$ in order to resolve the
deconfining phase transition at different vortex density cutoffs
$d=4,6,8,10$ and $12$, with $L_{max}=1.7$ and $L_{min}=0.3$.
In Fig.~\ref{fig:d4} we show the results extracted for $d=4$, namely,
the cluster size histogram, the potential $V(R)$ between the quark
and anti-quark and the spatial and temporal string tensions as well as the
maximal cluster fraction as a function of temperature. For all other densities
($d=6\ldots 12$) we show the cluster size histograms in Fig.~\ref{fig:hsT}
and string tensions resp. maximal cluster fractions vs. temperature in
Fig.~\ref{fig:sigT}. In the $d=4$ case (Fig.~\ref{fig:d4}) we observe a
phase transition in the vicinity of the inverse temperatures $L_T=5,6$. The
cluster size histogram in Fig.~\ref{fig:d4}a shows no cluster percolating
through the physical volume for $L_T=4$, whereas from $L_T=7$ onwards
one large percolating cluster starts to dominate the configuration.
The quark--anti-quark potential shown in Fig.~\ref{fig:d4}b is still flat
(asymptotically) for $L_T=5$, while a linearly rising behavior is evident
by $L_T=9$. In between, the potentials do not show a clear linear
behavior, and the determination of the string tension is somewhat ambiguous;
however, a deviation of the potential below an exactly linear behavior
is to be expected in view of the finite spatial extent of the physical
volume and the periodic boundary conditions. In the $d=4$ case, thus, the
deconfining transition is not very sharply defined; this is associated
with the rather small density cutoff $d$, as is revealed by examination
of higher values of $d$.

\begin{figure}[h]
	\centering
	a)\includegraphics[width=.48\linewidth]{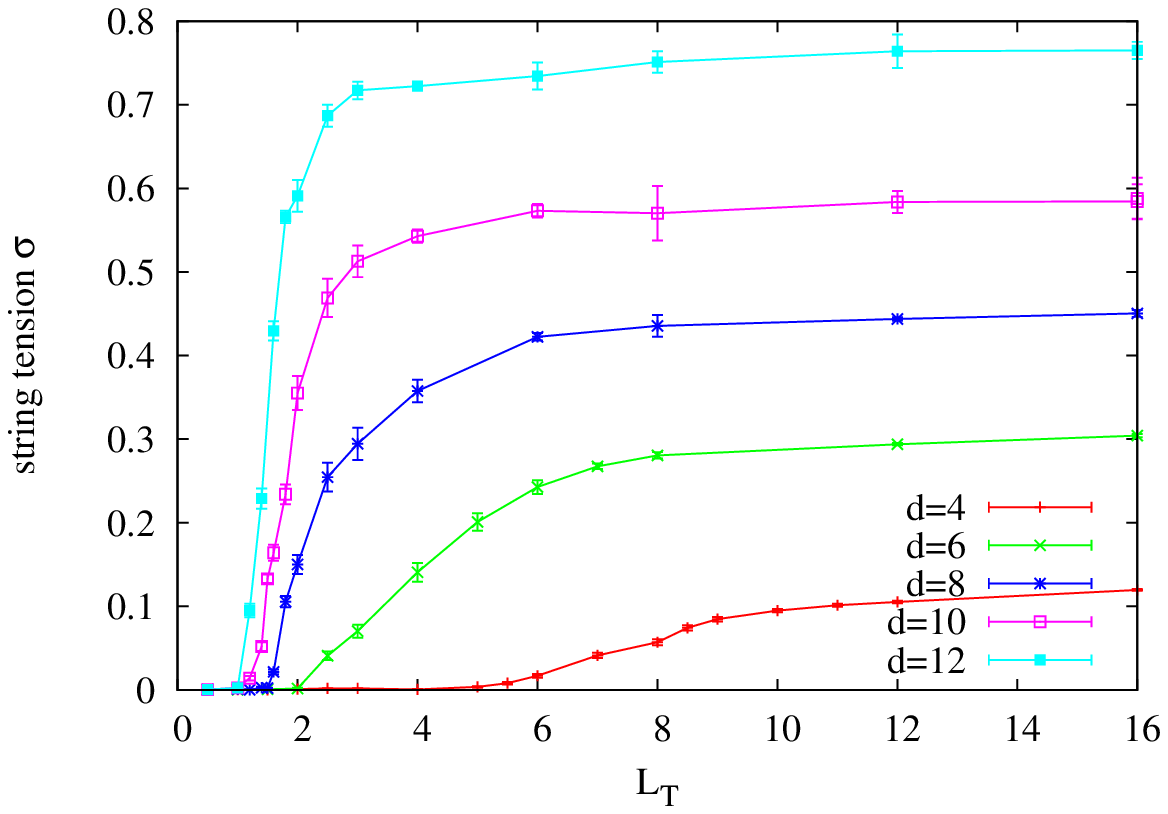}
	b)\includegraphics[width=.48\linewidth]{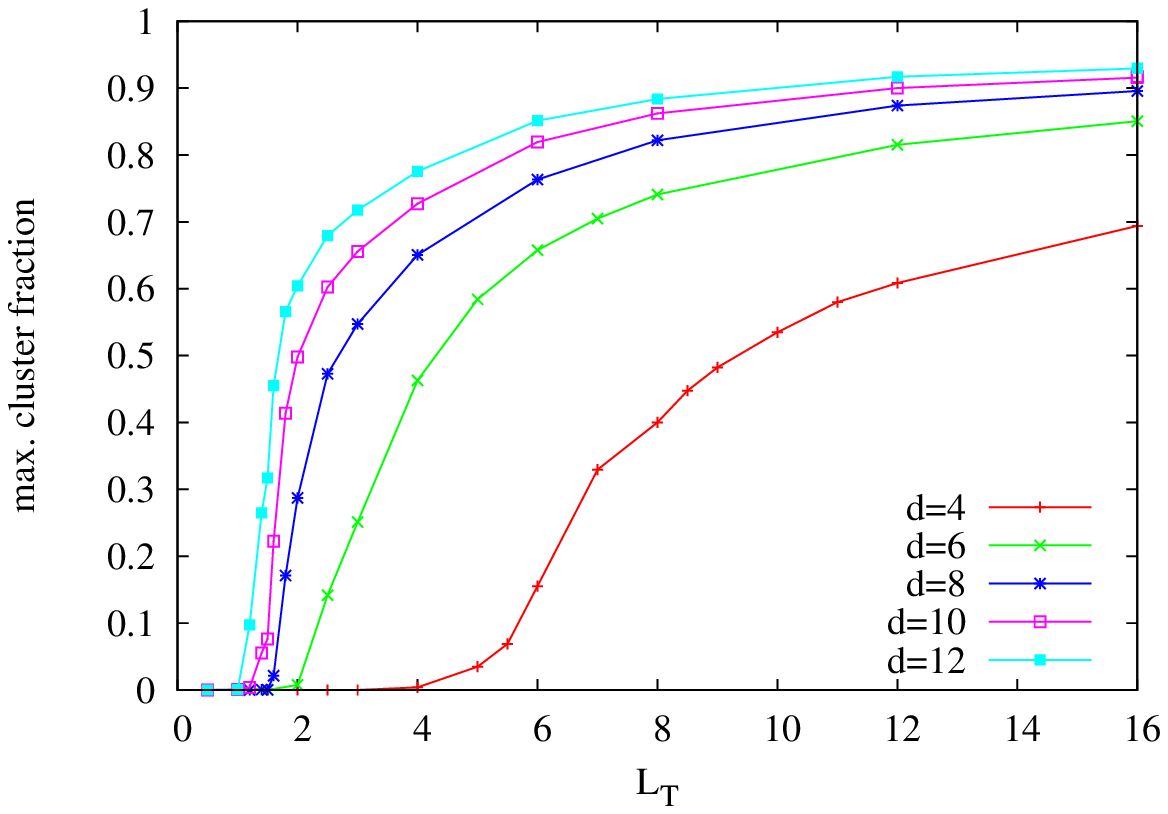}
	\caption{String tension $\sigma$ and maximal cluster fraction vs. time extent
	$L_T$ of $16^2\times L_T$ volumes, for different density cutoffs, vortex lengths $L_{max}=1.7, L_{min}=0.3$.}
	\label{fig:sumT}
\end{figure}

\begin{figure}[h]
	\centering
	a)\includegraphics[width=.48\linewidth]{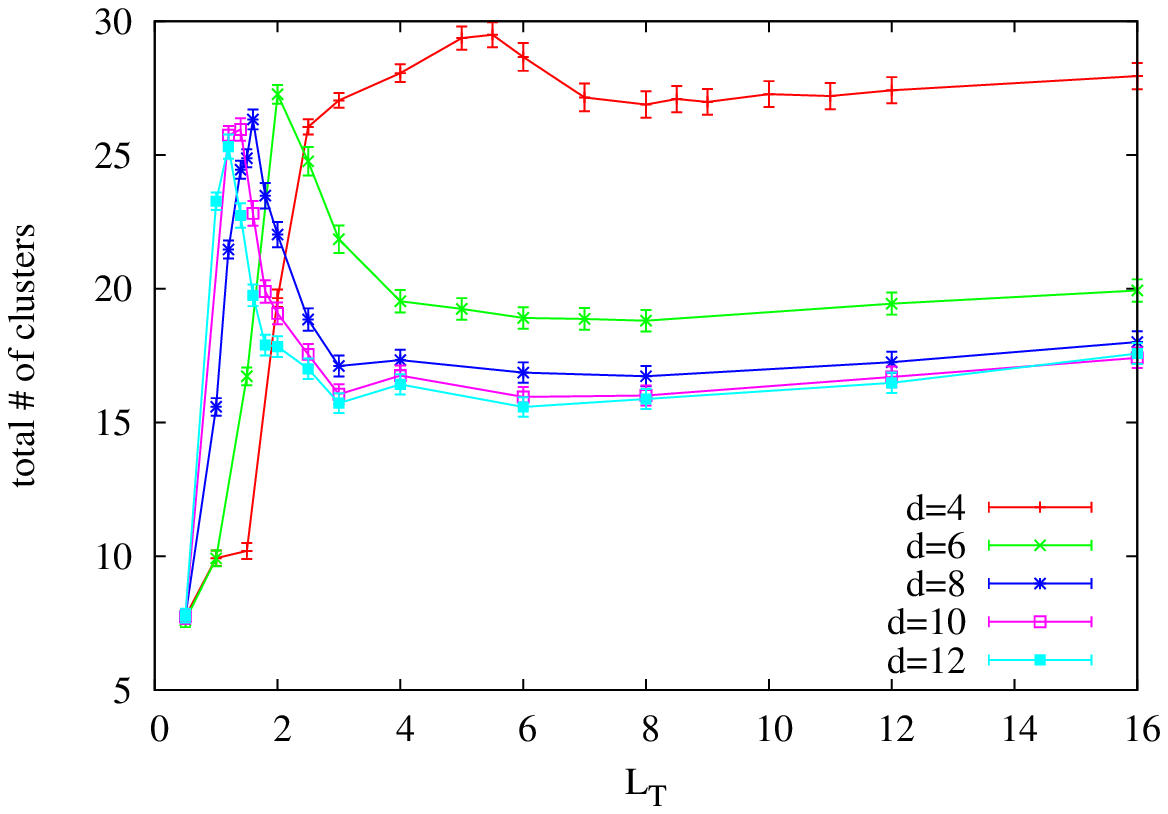}
	b)\includegraphics[width=.48\linewidth]{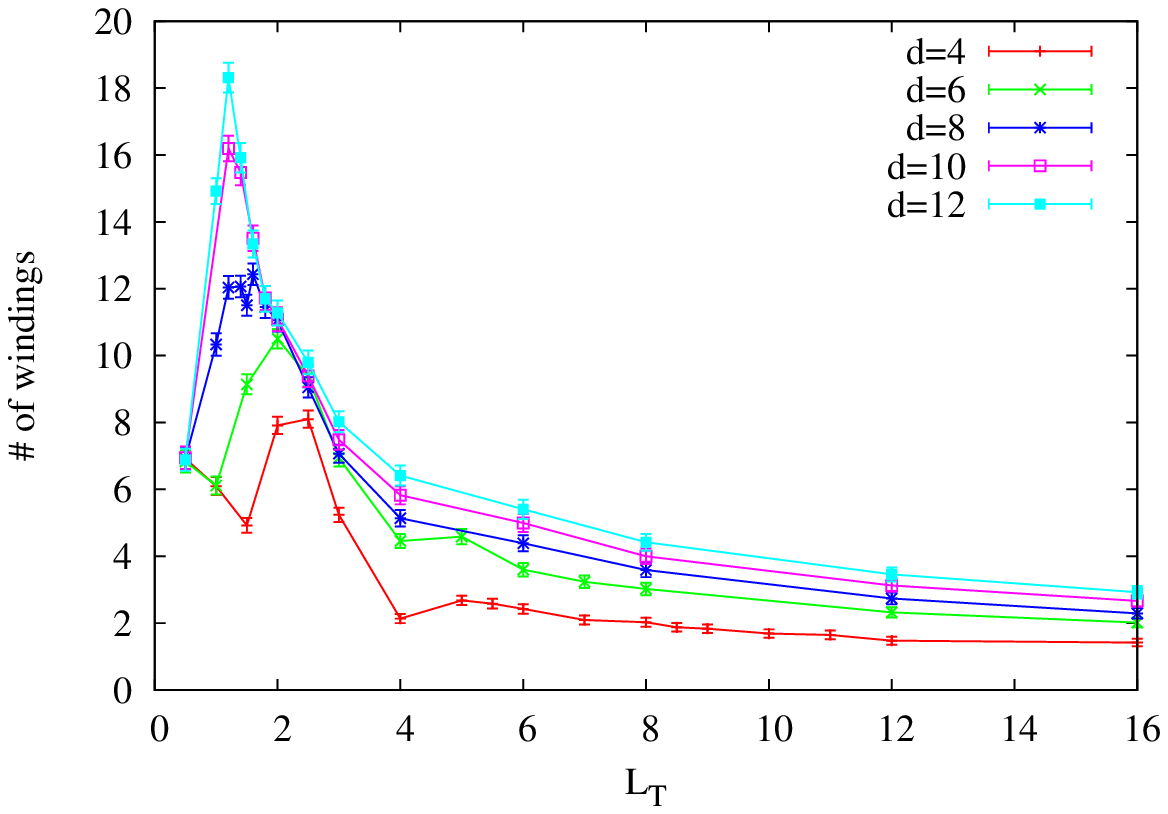}
	\caption{a) Total number of vortex clusters and b) windings
	around the temporal dimension, on $16\times L_T$ volumes, for vortex lengths $L_{max}=1.7, L_{min}=0.3$.}
	\label{fig:clusT}
\end{figure}

\begin{figure}[p]
	\centering
	a)\includegraphics[width=.33\linewidth]{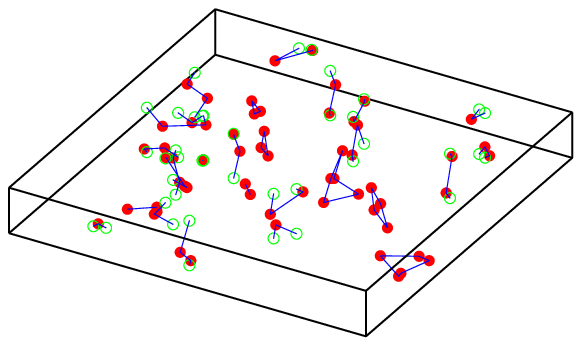}
	\hspace{1cm}\includegraphics[width=.33\linewidth]{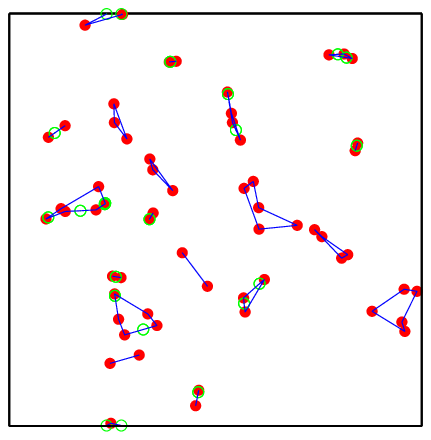}
	\vspace{-1cm}\\
	b)\includegraphics[width=.33\linewidth]{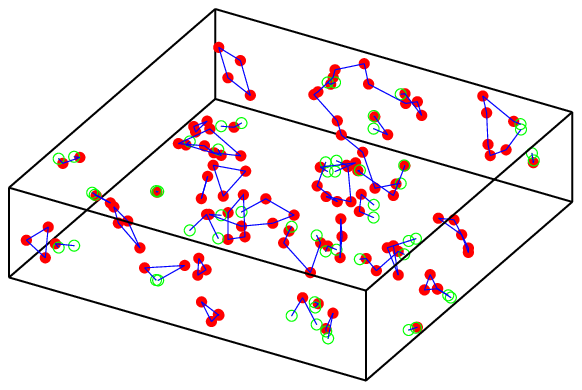}
	\hspace{1cm}\includegraphics[width=.33\linewidth]{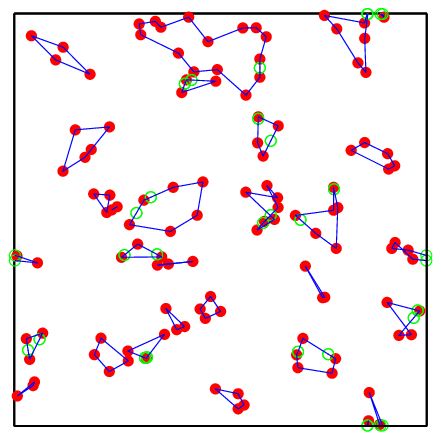}
	\vspace{-1cm}\\
	c)\includegraphics[width=.33\linewidth]{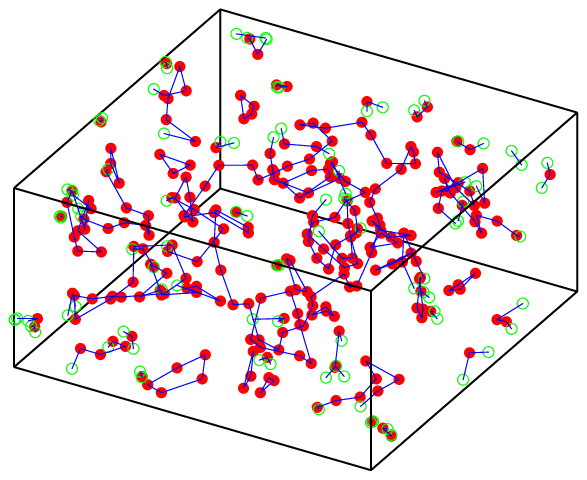}
	\hspace{1cm}\includegraphics[width=.33\linewidth]{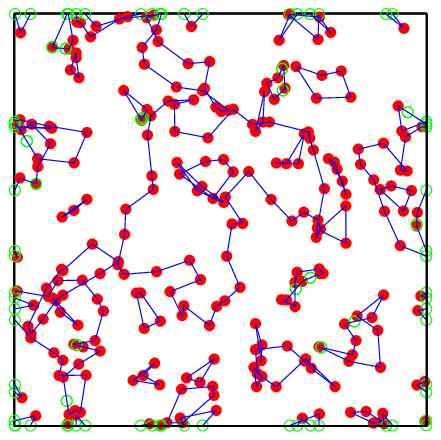}
	\vspace{-2cm}\\
	d)\includegraphics[width=.4\linewidth]{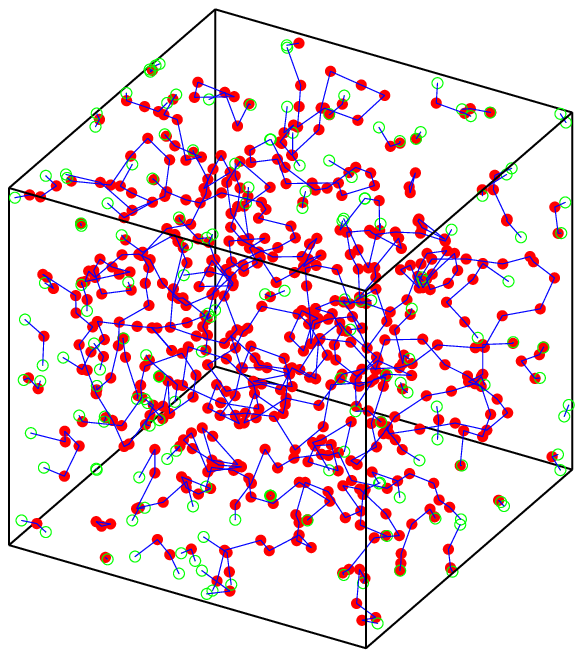}
	\includegraphics[width=.33\linewidth]{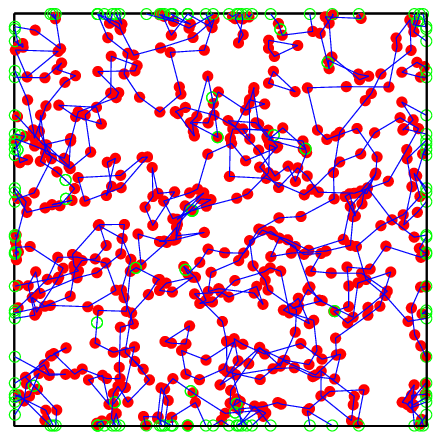}
	\caption{$16^2\times$ a) $2$, b) $4$, c) $8$ and d)
	$16$ sample configurations, density cutoff $d=4$, vortex lengths
$L_{max}=1.7, L_{min}=0.3$. The right-hand views are directly into the
temporal direction.}
	\label{fig:conf3D}
\end{figure}

For higher vortex densities, the phase transition becomes much sharper
and the inverse critical temperature tends to smaller temporal
extents $L_T$. This can be seen in Fig.~\ref{fig:sigT} and especially
Fig.~\ref{fig:sumT}, which summarizes the results on the finite temperature
phase transition for various vortex density cutoffs $d$, and also in the
corresponding cluster size histograms in  Fig.~\ref{fig:hsT} where one should
note the different coloring and temperatures for the individual plots. We
locate the phase transition for $d=6$ in the vicinity of $L_T=2$, for $d=8$
in the vicinity of $L_T=1.6$, for $d=10$ in the vicinity of $L_T=1.5$ and
for $d=12$ in the vicinity of $L_T=1.2$. Further, we notice that
Fig.~\ref{fig:d4}c and the plots in Fig.~\ref{fig:sigT} show a perfect
agreement between the confinement (string tension) and percolation
(maximal cluster fraction) transitions. In Fig.~\ref{fig:conf3D} we
show sample configurations for various temperatures and density cutoff
$d=4$. While for $L_T=2$ and $4$ (Fig.~\ref{fig:conf3D}a and b) we see
many small vortex clusters, we observe already one big cluster extending
over the whole physical volume together with some small clusters for $L_T=8$
(Fig.~\ref{fig:conf3D}c) while for $L_T=16$ (zero temperature,
Fig.~\ref{fig:conf3D}d) it appears as though almost all nodes were
connected. In fact, the careful observer can still make out a few
three- and four-node clusters, {\it e.g.}, at the bottom left corner
of the 3D plot in Fig.~\ref{fig:conf3D}d, and indeed we still have
around 25-30 individual clusters in this configuration, see
Fig.~\ref{fig:clusT} for the average number of clusters within a
configuration at different inverse temperatures $L_T$ and vortex
densities $d$. Nevertheless, the majority ($\approx 70\%$,
see Fig.~\ref{fig:d4}a and c) of nodes in Fig.~\ref{fig:conf3D}d
is part of one big cluster percolating through the whole physical
volume, indicating a confined phase.

It remains to discuss the spatial string tension, which at first sight
seems to display unusual behavior in Figs.~\ref{fig:d4}c
and~\ref{fig:sigT}. Apart from the fact that the behavior at very small
$L_T$ becomes unphysical, because the lower bound on the vortex segment
length $L_{min} $ artificially suppresses vortices winding around the time
direction, one would expect that the spatial string tension remains more
or less constant across the deconfining transition; after all, spatial
Wilson loops will still be pierced by vortex clusters winding in the time
direction even once the percolation effect ceases. However, it appears
that these two effects to a certain extent disentangle and are separated
as a function of $L_T$. For $d=4$ we clearly see a decreasing spatial
string tension with increasing temperature, in accordance with loss of
percolation in the vicinity of the percolation transition, while only below
$L_T=4$, ({\it i.e.}, above the transition temperature) the effect of winding
vortices sets in. Note that (much weaker) hints of such
behavior are also seen in vortex ensembles extracted from lattice
Yang-Mills configurations~\cite{Langfeld:1998cz}. In Fig.~\ref{fig:clusT}b
we plot the number of vortices winding around the time direction;
the correlation of these windings with the behavior seen in the spatial
string tension for $d=4$ is clearly visible. For higher densities,
percolation and winding effects become more entangled and harder to
distinguish.

\clearpage

Examining once more Fig.~\ref{fig:clusT}, we observe that the total number
of vortex clusters peaks around the deconfinement temperature in all cases.
This peak is rather weak for $d=4$, but becomes stronger as $d$ is allowed
to rise. The behavior of the plots in Fig.~\ref{fig:clusT} converges with
rising $d$; only the $d=4$ case is fairly strongly separated from the ones
at higher $d$. By $d=8$, the behavior of the vortex configurations appears
to have converged, and the finite temperature transition seen in
Fig.~\ref{fig:sumT} has become sharp. For this reason, we choose $d=8$
for the analysis of the dependence on vortex segment length range
in Secs.~\ref{sec:lmax} and~\ref{sec:lmin}.

\subsection{Deconfining transition as a function of vortex density cutoff $d$}

\begin{figure}[h]
	\centering
	a)\includegraphics[width=.48\linewidth]{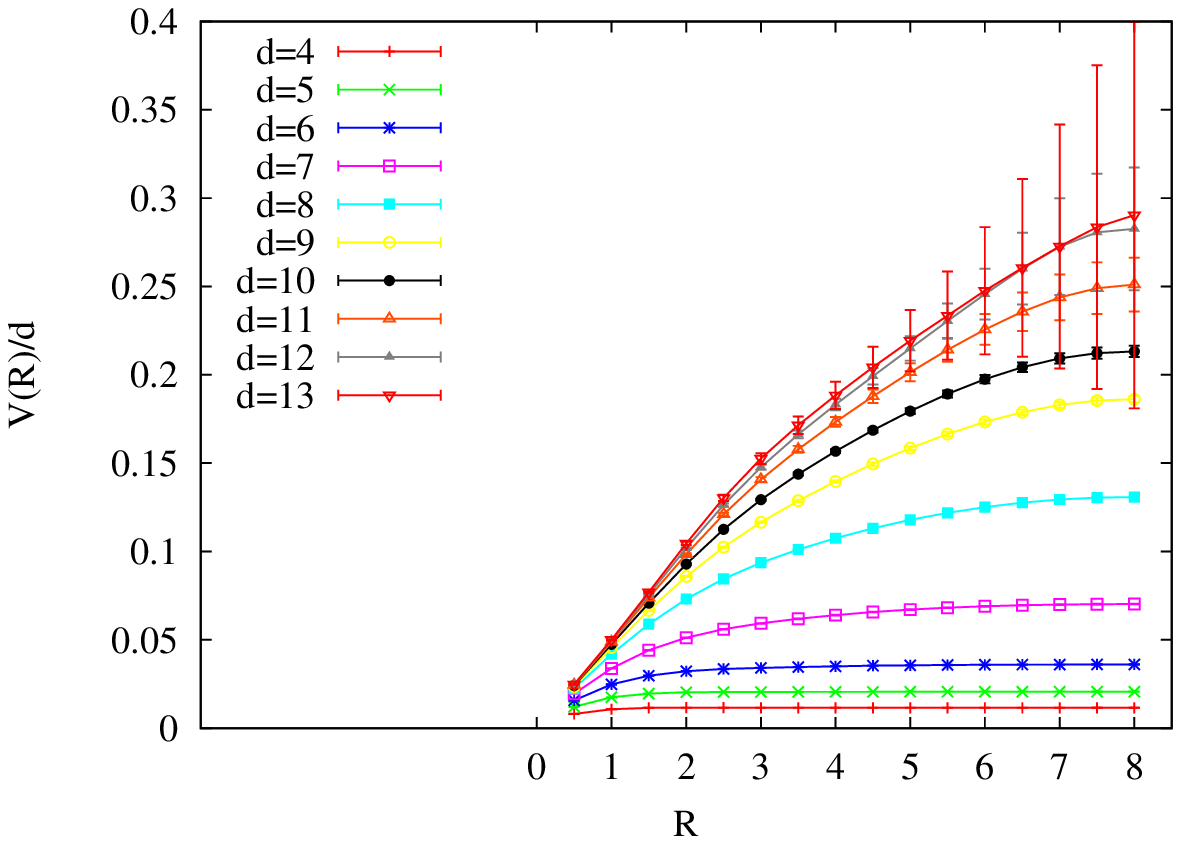}
	b)\includegraphics[width=.48\linewidth]{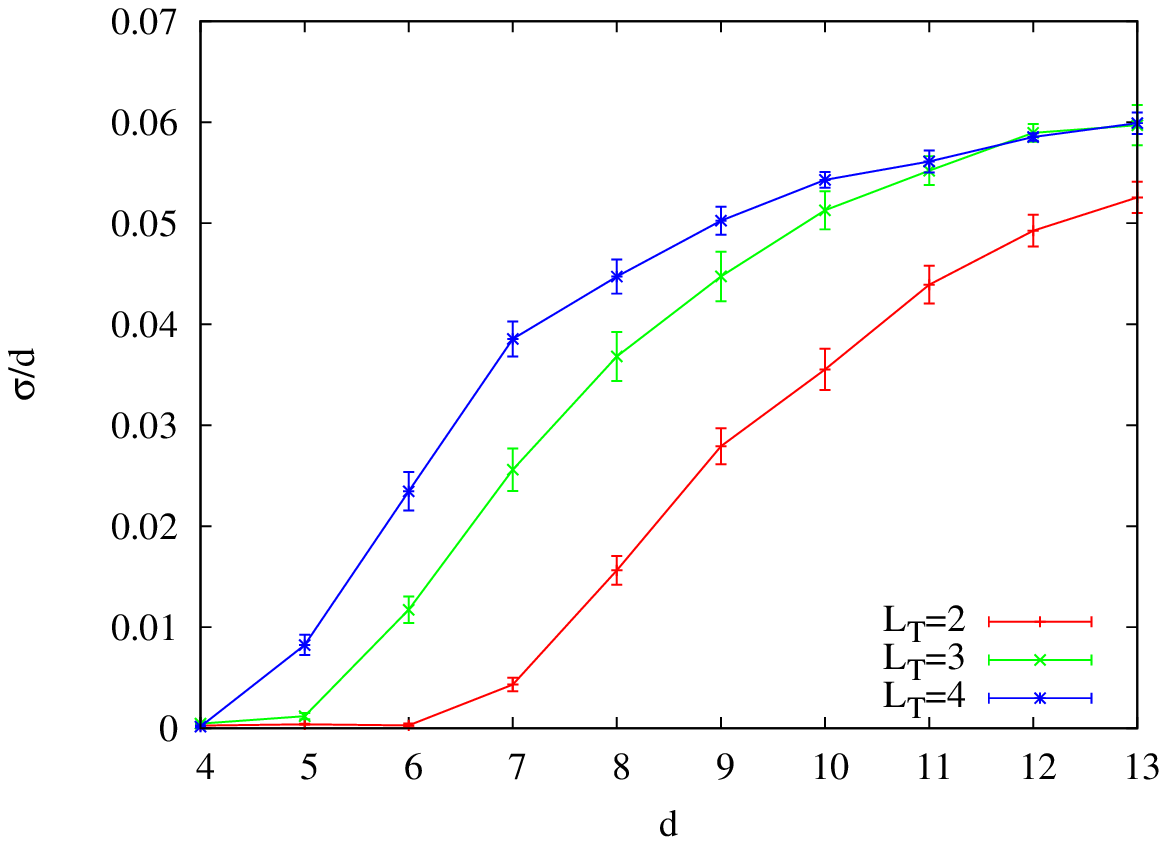}
	\vspace{2mm}\\
	c)\includegraphics[width=.48\linewidth]{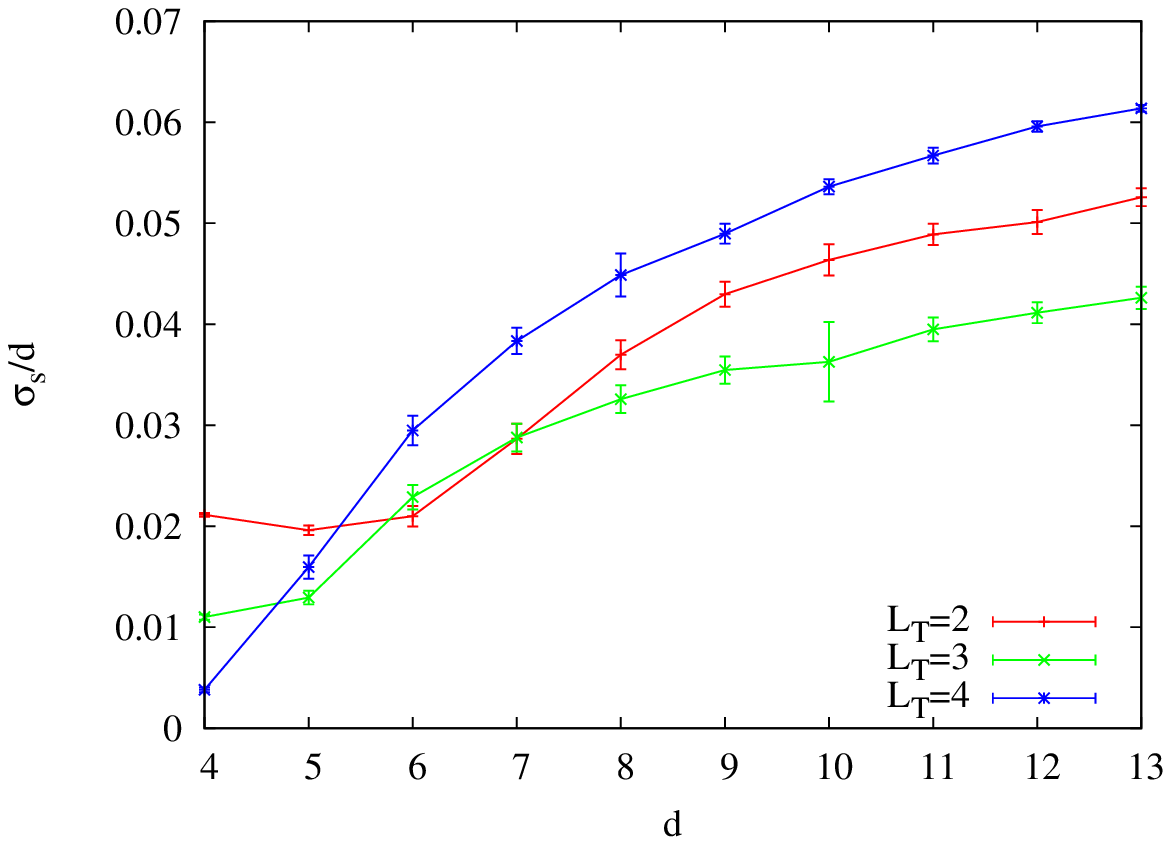}
	d)\includegraphics[width=.48\linewidth]{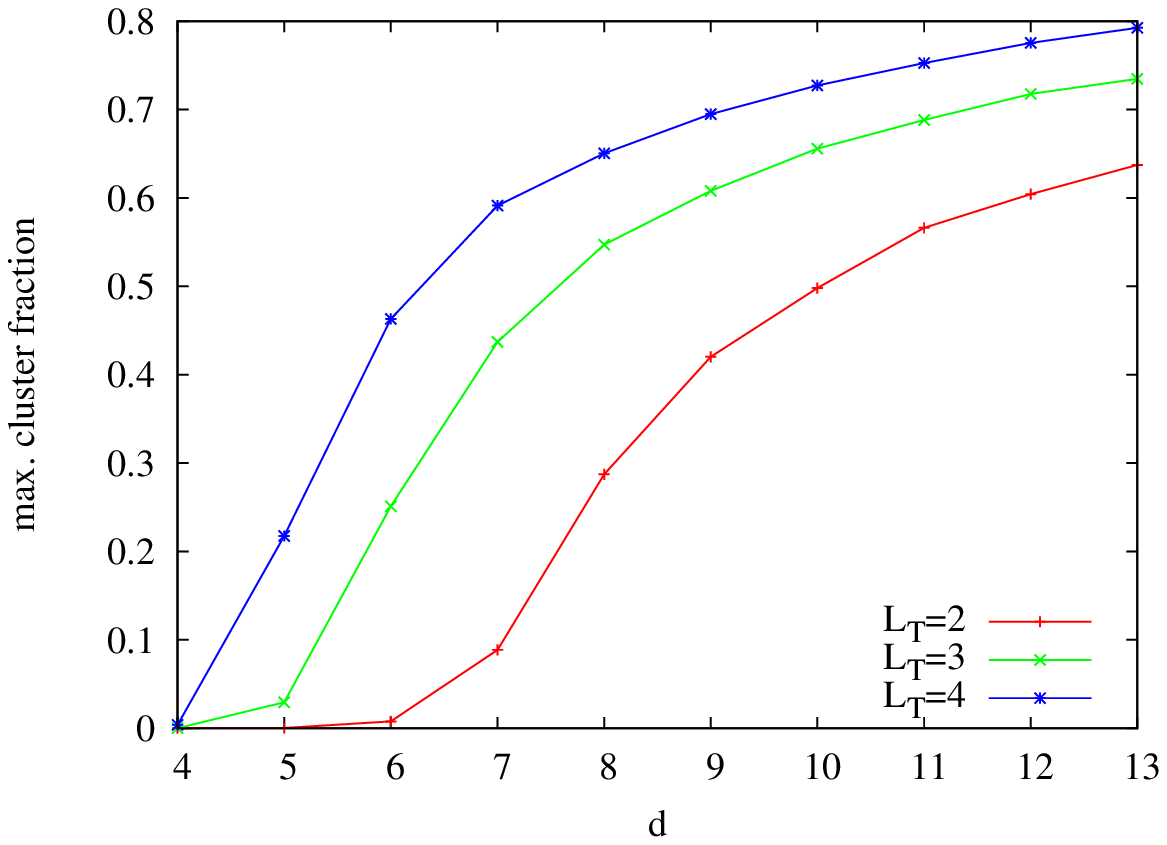}
	\caption{a) Quark--anti-quark potentials on a $16^2 \times2$ volume, string tensions b)
	$\sigma$, c) $\sigma_s$ and d) maximal vortex cluster fraction as a function of density cutoff $d$ for $16^2\times
	L_T$ volumes and vortex lengths $L_{max}=1.7, L_{min}=0.3$.}
	\label{fig:sigd}
\end{figure}

\begin{figure}[h]
	\centering
	a)\includegraphics[width=\linewidth]{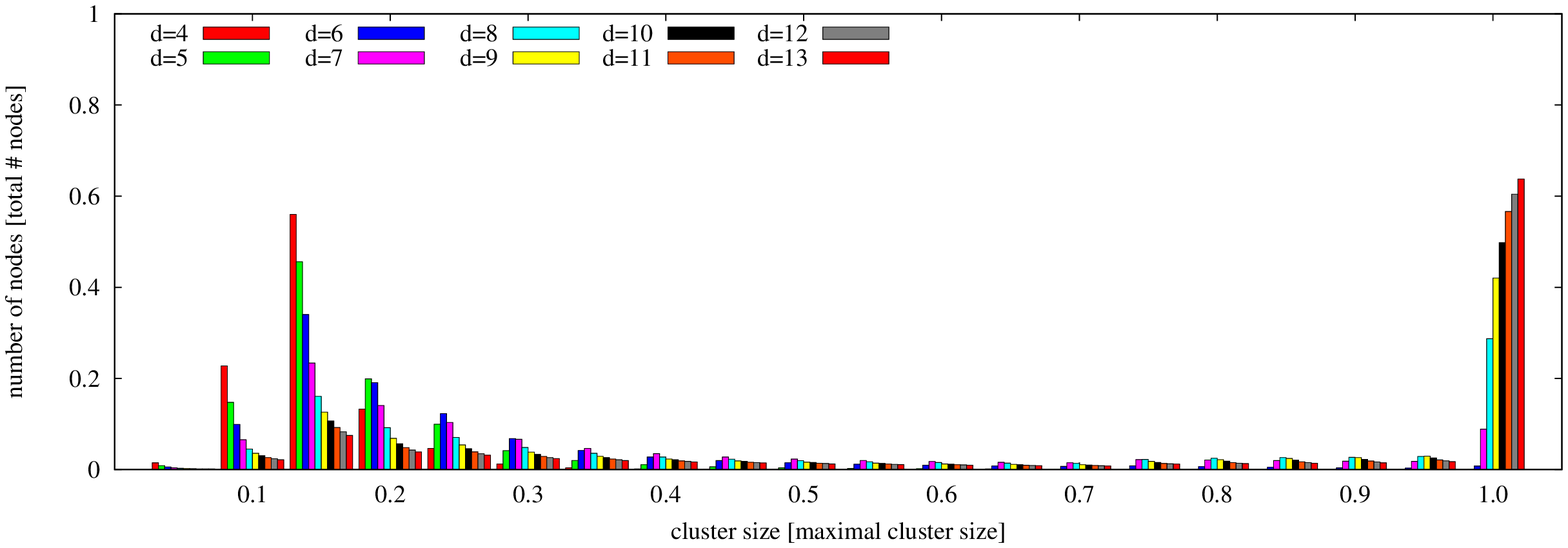}
	\vspace{1mm}\\
	b)\includegraphics[width=\linewidth]{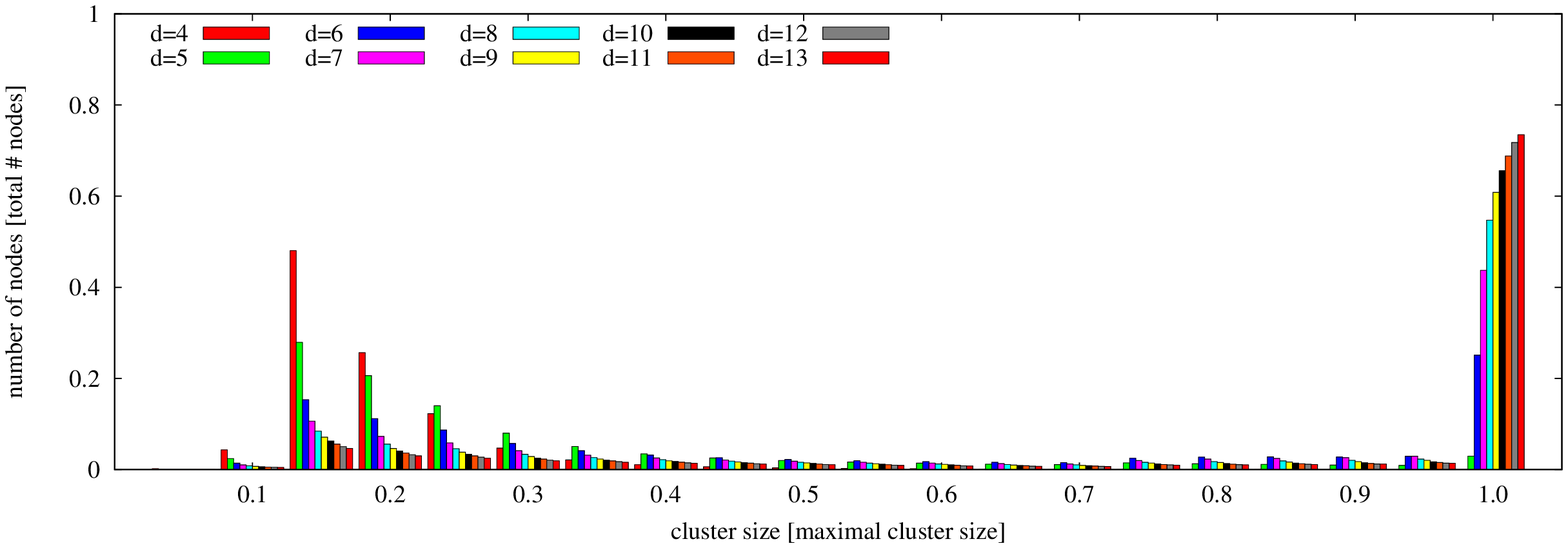}
	\vspace{1mm}\\
	c)\includegraphics[width=\linewidth]{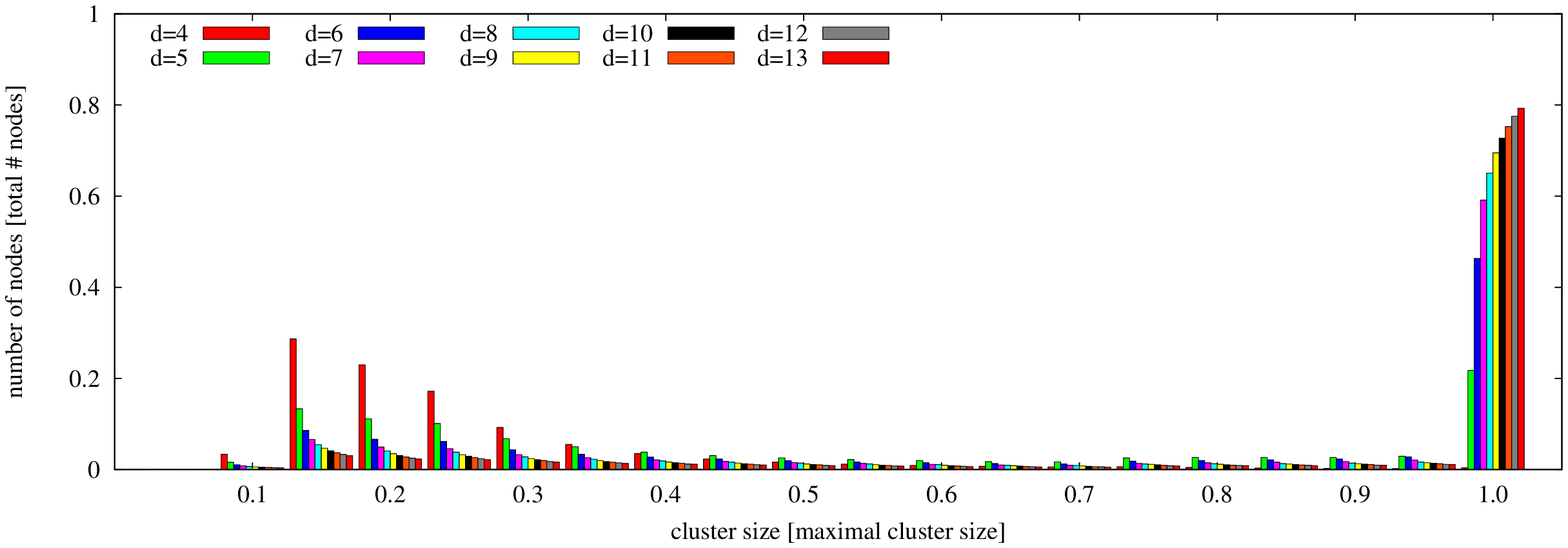}
\caption{Cluster size histograms for a) $16\times 2$ b) $16\times 3$ c)
$16\times 4$ volumes and different density cutoffs $d$, vortex lengths $L_{max}=1.7, L_{min}=0.3$.}
	\label{fig:hstd}
\end{figure}

In this section, we investigate the random vortex line ensembles for
varying density cutoff $d=4\ldots 13$ on three different physical volumes
$16^2\times L_T$ with $L_T=2\ldots 4$, $L_{max}=1.7$ and $L_{min}=0.3$. In
Fig.~\ref{fig:sigd} we show the quark--anti-quark potentials on a
$16^2 \times2$ volume, as well as the string tensions $\sigma$ and
$\sigma_s$ and the maximal vortex cluster fraction as a function of $d$
for the various $L_T $. In Fig.~\ref{fig:hstd} we show the corresponding
cluster size histograms. We observe a deconfinement transition with
respect to the vortex density cutoff $d$. At $d=4$ all cases are in the
deconfined phase; the $L_T=4$ configurations then immediately start to
confine when $d$ is increased, whereas the $L_T=3$ and $L_T=2$ configurations
reach the transition around $d=5$ and $d=6$, respectively. Again, the 
confinement and percolation transitions (Fig.~\ref{fig:sigd}b and d)
agree perfectly; the maximal clusters in Fig.~\ref{fig:hstd} start to
develop at exactly the aforementioned critical densities. Higher densities
of course allow for more reconnections and percolation, {\it i.e.}, they
facilitate confinement. The spatial string tension at $d=4$ essentially
vanishes in the $L_T=4$ case with percolation having ceased and almost
no winding vortex clusters present to counteract the decline; on the other
hand, for $L_T=3$ and $L_T=2$, the effect of winding vortices already
manifests itself at $d=4$ in a stabilization of the spatial string tension
at finite values. This interplay was discussed in more detail already in
the previous section.

\subsection{Phase transition from varying maximal vortex length $L_{max}$}\label{sec:lmax}

\begin{figure}[h]
	\centering
	a)\includegraphics[width=\linewidth]{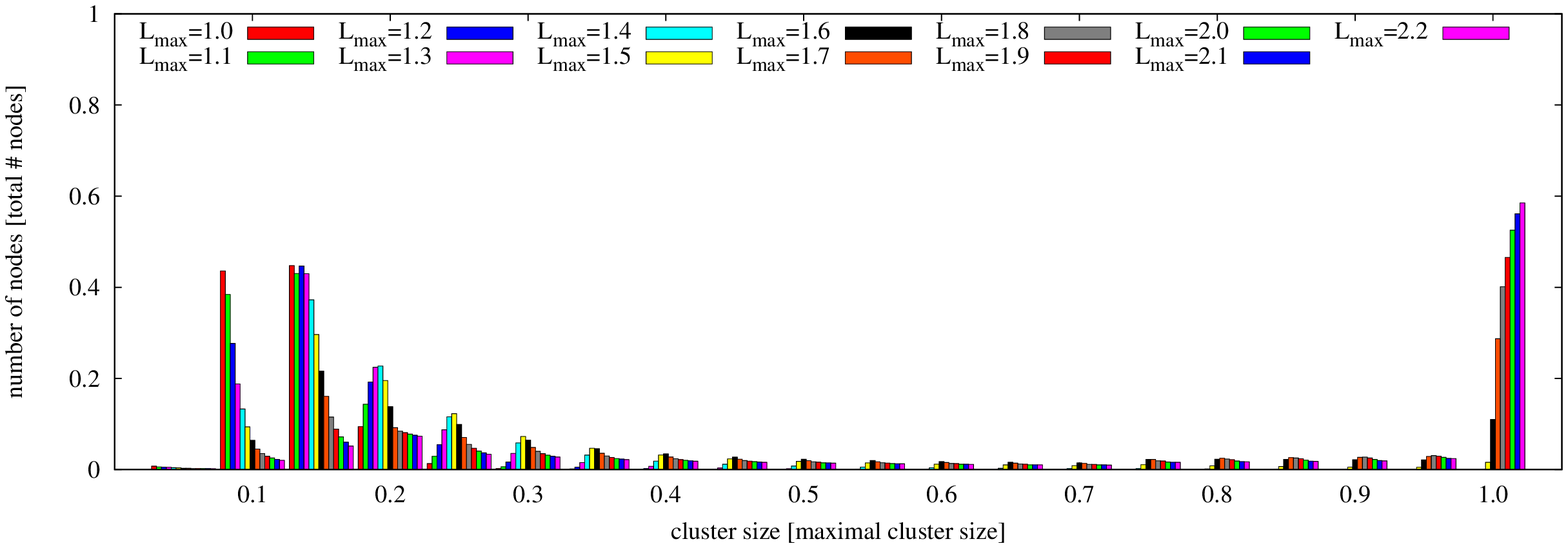}
	b)\includegraphics[width=.48\linewidth]{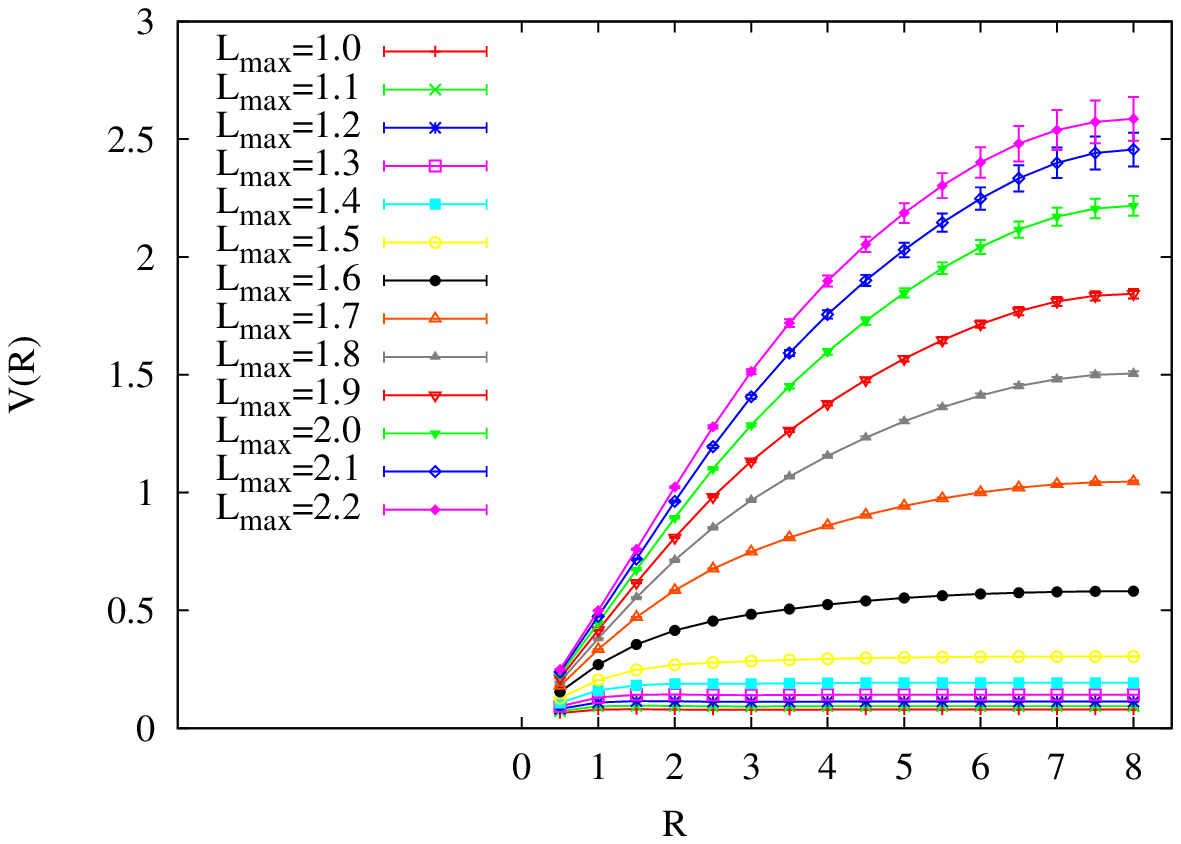}
	c)\includegraphics[width=.48\linewidth]{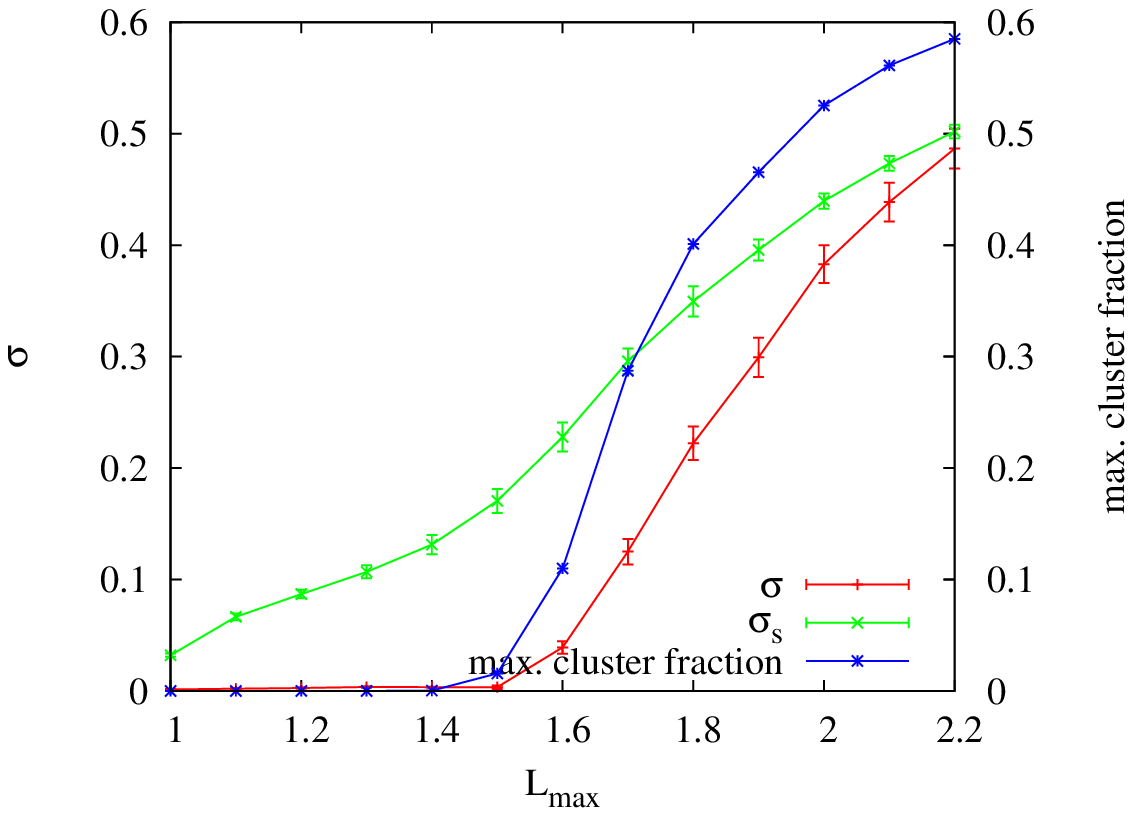}
	\caption{a) Cluster size histogram, b) quark--anti-quark potential and c)
	string tensions $\sigma$ and $\sigma_s$ as well as maximal cluster fraction as a function of maximal vortex segment length $L_{max}$, for a $16\times2$
	volume, density cutoff $d=8$ and $L_{min}=0.3$.}
	\label{fig:lmax}
\end{figure}

In this section, we investigate the behavior of the ensembles at different
maximal vortex segment lengths $L_{max}=1.0\ldots 2.2$ for a physical volume
$16^2\times2$ and a density cutoff $d=8$, with $L_{min}=0.3$. 
In Fig.~\ref{fig:lmax} we show the cluster size histogram, the
quark--anti-quark potential, and string tensions $\sigma$ and $\sigma_s$
as well as maximal cluster fraction versus the different maximal vortex
segment lengths $L_{max}$. 

In this case, we observe a well-defined phase transition at $L_{max}=1.5$,
with no percolation and string tension $\sigma$, {\it i.e.}, a flat
quark--anti-quark potential below that threshold, and percolation
resp. confinement above. Restricting the vortex line segment length
to a more stringent upper bound shifts the action-entropy balance away
from the entropy-dominated regime and leads to small, separated
vortex clusters which cannot reconnect or percolate, and confinement is lost.
Evidently, also the spatial string tension in the deconfined phase
decreases as the $L_{max} $ bound becomes more stringent, indicating that
the number of vortices winding in the time direction is likewise suppressed.

\clearpage

\vspace{-5mm}

\subsection{Behavior as a function of minimal vortex/reconnection length $L_{min}$}\label{sec:lmin}

\begin{figure}[h]
	\centering
	a)\includegraphics[width=\linewidth]{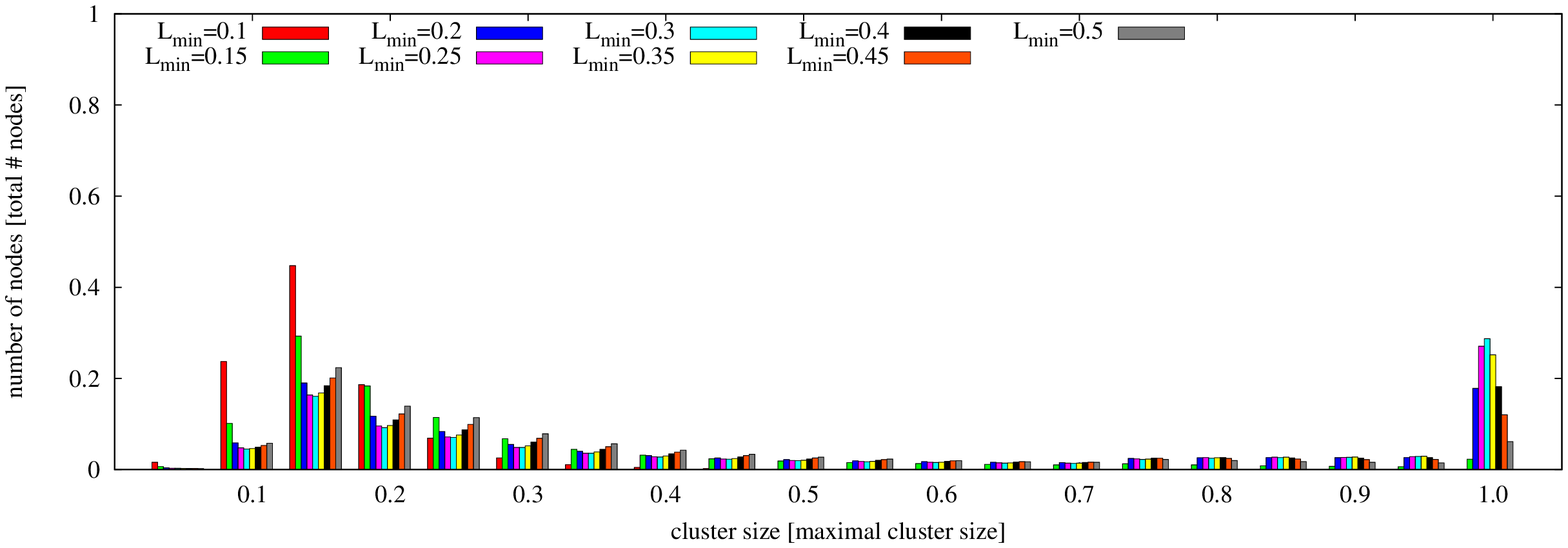}
	b)\includegraphics[width=.48\linewidth]{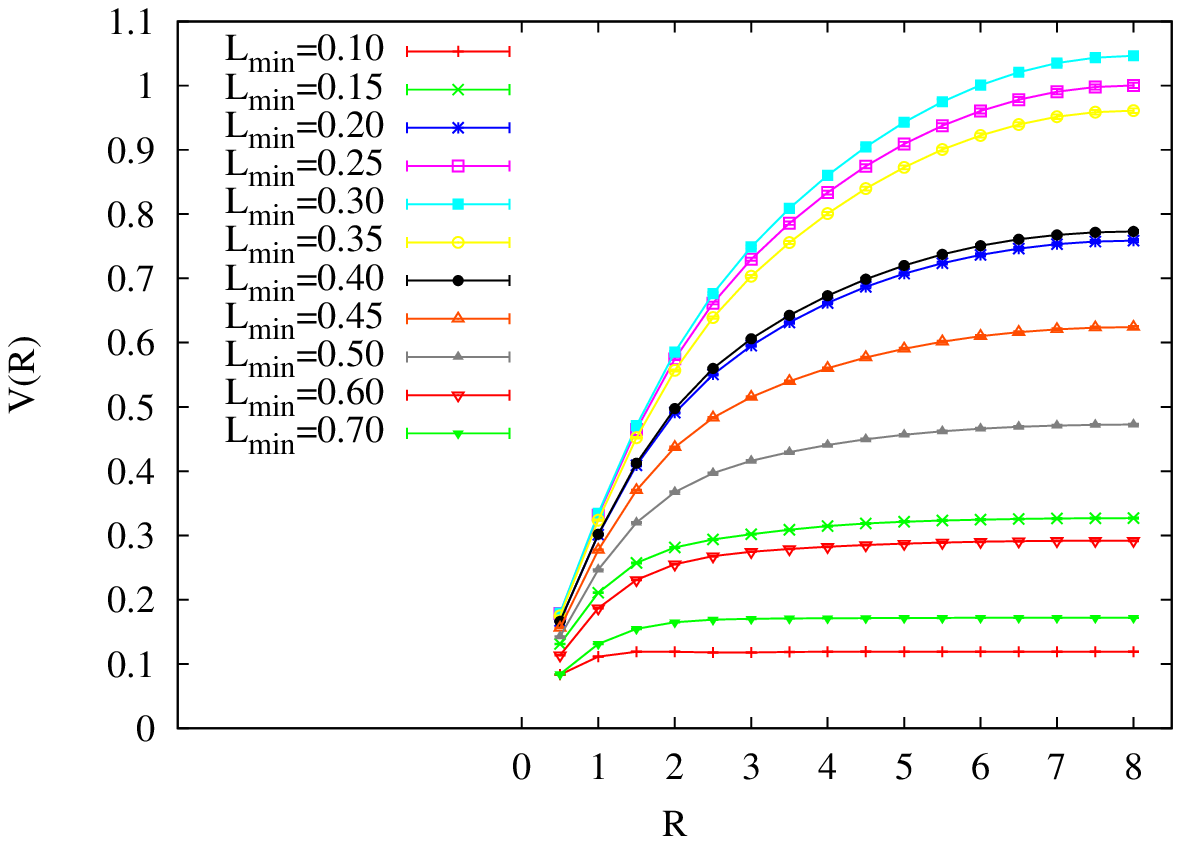}
	c)\includegraphics[width=.48\linewidth]{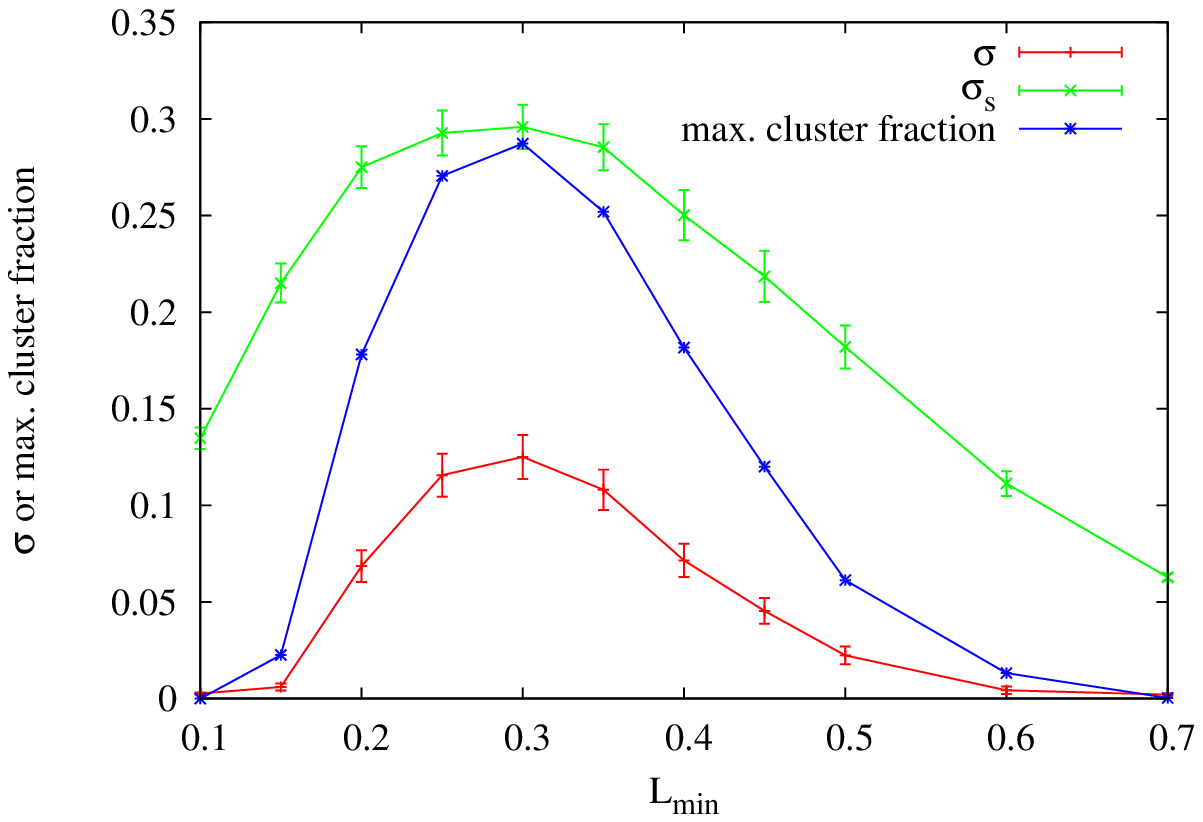}
	d)\includegraphics[width=.48\linewidth]{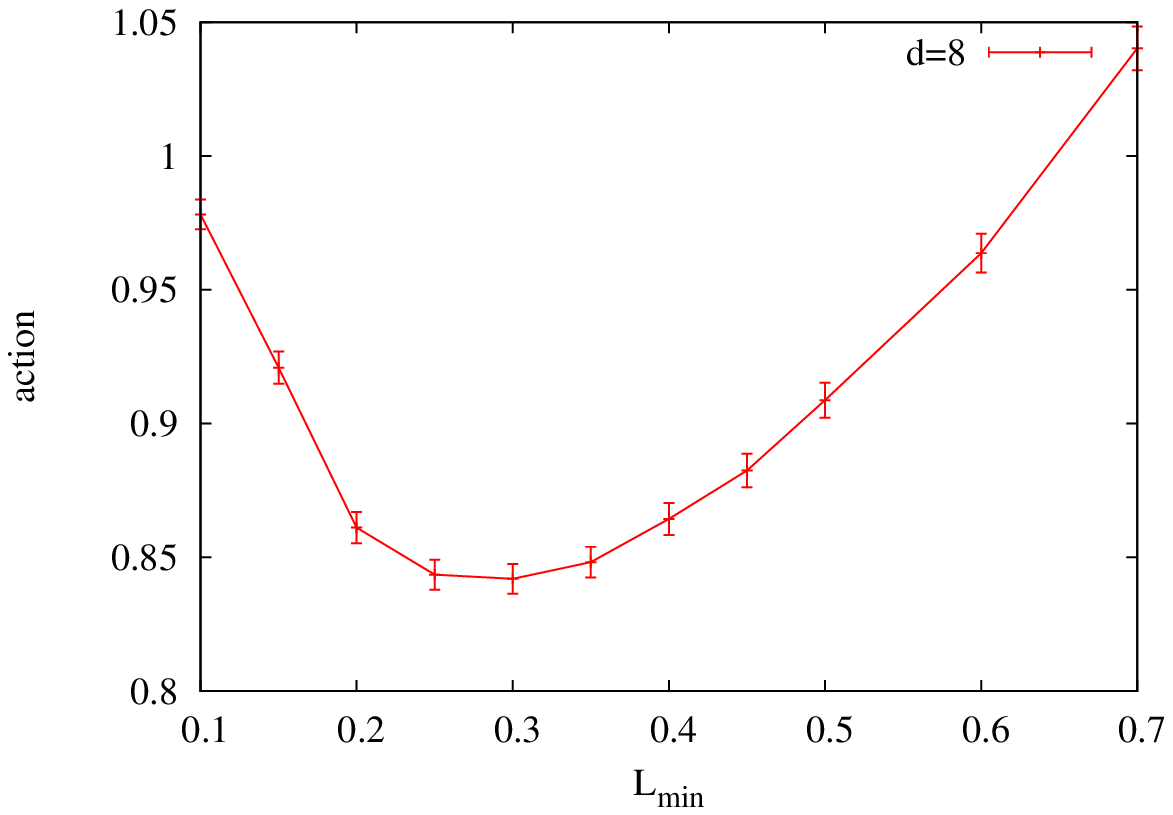}
	e)\includegraphics[width=.48\linewidth]{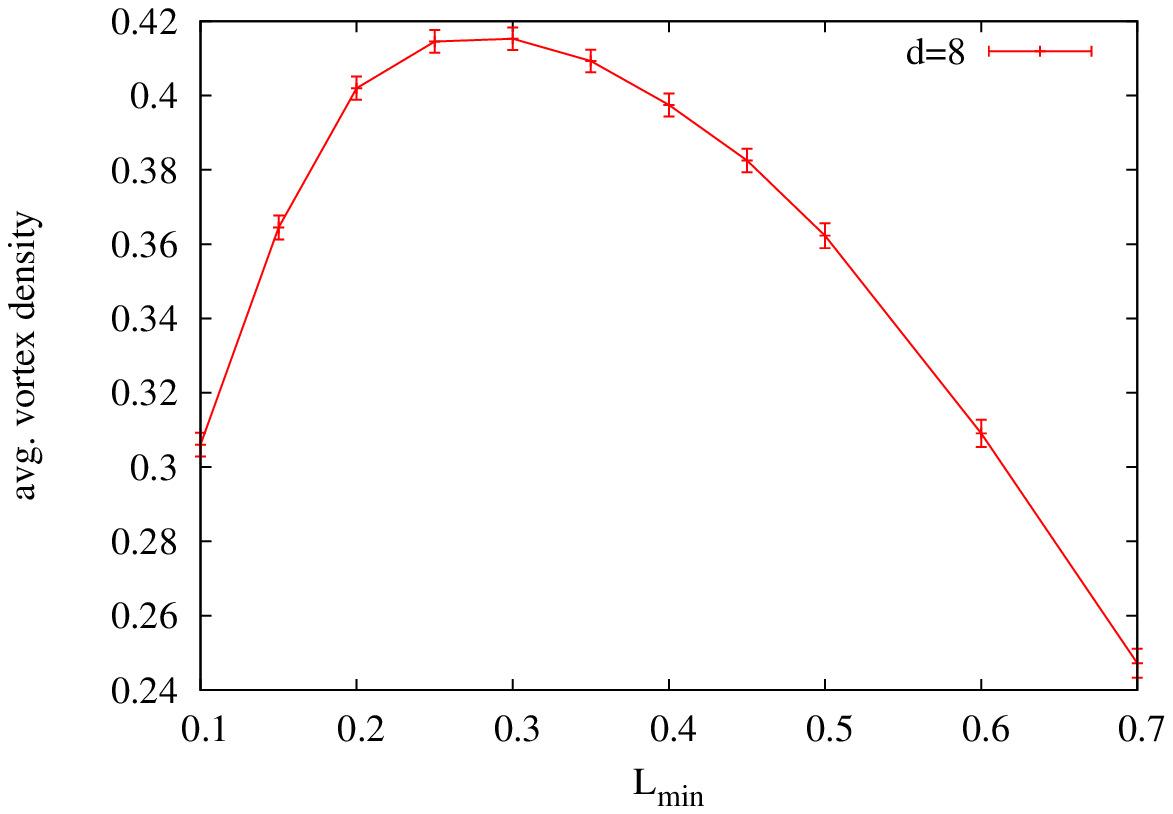}
	\caption{a) Cluster size histogram, b) quark--anti-quark potential, c) string
		tensions and maximal vortex cluster fraction, average d) node action and e)
	vortex line density (= avg. node density $\times$ avg. vortex length) for $16\times2$ volumes, density cutoff $d=8$ and different vortex/reconnection lengths $L_{min}$ at $L_{max}=1.7$.}
	\label{fig:lmin}
\end{figure}

The interpretation of the phase space with respect to the minimal vortex
segment length $L_{min}$ is more complex than with respect to other
parameters. $L_{min}$ defines a minimal length scale which enters a
number of effects governing the vortices; it not only restricts the minimal
length of a vortex segment itself, but also determines the maximal move
radius $r_m=4L_{min}$, the maximal add radius $r_a=3L_{min}$ and the
reconnection distance $r_r=L_{min}$. That means that if we choose a small
$L_{min}$, the vortex clusters will not spread out very quickly
and reconnections are strongly suppressed. On the other hand, a large
$L_{min}$ restricts the set of available configurations, and thus drives the
system away from the entropy-dominated regime, while at the same time
obstructing equilibration, with large attempted updates and frequent
recombinations. Both limits do not realize
the physical behavior we want to study, and our analysis of configurations in
$16^2\times2$ volumes with vortex density $d=8$, maximal vortex length
$L_{max}=1.7$ and varying $L_{min}$ seems to confirm these expectations.
The aforementioned set of fixed parameters lies close to the critical point
for all the deconfining phase transitions studied further above,
{\it i.e.}, as a function of temperature, vortex density and maximal
vortex segment length $L_{max}$. In Fig.~\ref{fig:lmin} we plot the cluster
size histogram, the quark--anti-quark potentials, string tensions $\sigma$,
$\sigma_s$ and maximal cluster fraction as well as average vortex node
action and vortex density versus $L_{min}=0.1\ldots 0.7$ in steps of $0.05$.
We observe deconfined phases for both very small and very large $L_{min}$.
In the former case, the configurations remain rather static and do not
readily recombine and percolate; in the latter case, the space of
configurations is restricted, leading to a suppressed vortex density
which also does not exhibit good percolation properties. At $L_{min}=0.3$,
however, we find a common maximum for string tensions, maximal cluster
fraction and average vortex density, and simultaneously the average action
shows a minimum. This validates our initial choice for $L_{min}=0.3$,
which appears to yield rather stable ensembles that permitted reliable
studies of the phase transitions investigated further above.

\clearpage

\vspace{-5mm}

\section{Conclusions \& Outlook}\label{sec:concl}

We presented a $2+1$-dimensional center vortex model of the Yang-Mills vacuum.
The vortices are represented by closed random lines which are modeled as being
piece-wise linear, and an ensemble is generated by Monte Carlo methods. The
physical space in which the vortex lines are defined is a torus with periodic
boundary conditions. The motivation for this study was to explore a
formulation which avoids the shortcomings of previous realizations of
random center vortex models that relied on a hypercubic scaffolding for
the construction of the vortex configurations. The present formulation
preserves translational and rotational symmetry, and updates can occur
continuously in space-time. Vortex configurations are allowed to grow and
shrink, and also reconnections are allowed, {\it i.e.}, vortex lines may
fuse or disconnect. Our ensemble therefore contains not a fixed, but a
variable number of closed vortex lines. This is in fact a crucial
ingredient for achieving a system of percolating vortices, i.e., a
confining phase. All vortex updates (move, add, delete, reconnect) are
subjected to a Metropolis algorithm driven by an action depending on
vortex segment length and the angle between two adjacent vortex segments;
{\it i.e.}, the action contains both a length and a curvature term. After
tuning all necessary parameters, which are summarized in
section~\ref{sec:par}, we use the model to study both vortex
percolation and the potential $V(R)$ between quark and anti-quark
as a function of distance $R$ at different vortex densities, vortex
segment length ranges, reconnection conditions and at different
temperatures (by varying the temporal extent of the physical volume). 

We have found three deconfinement transitions, namely, as a function of
density, as a function of vortex segment length range, and as a function
of temperature. The deconfinement transitions coincide with percolation
transitions in the vortex configurations. For small vortex densities and
restricted vortex segment lengths, the configurations consist of small,
independent vortex clusters, and for high temperatures, vortex clusters
prefer to separate and wind around the (short) temporal extent of the
volume; in these cases, there is no percolation, the quark--anti-quark
potentials show no linearly rising behavior, {\it i.e.}, no string tension
is measured, and the system is in the deconfined phase. Once one allows for
higher vortex densities, less restricted vortex segment lengths, or larger
temporal extent, {\it i.e.}, lower temperature, the vortex configurations
begin to percolate; small clusters reconnect to mainly one large vortex
cluster filling the whole volume. In this regime, we measure a finite
string tension, {\it i.e.} linearly rising quark--anti-quark potentials;
hence the vortices confine quarks and anti-quarks. 

\clearpage

The physically most relevant extension of the modeling effort presented
here is of course the one to $D=4$ space-time dimensions, where center
vortices are described by 2-dimensional random world-surfaces. The
surfaces can be represented by random triangulations, anchored again
by nodes which can move, be added, or deleted from configurations.
Surface separation at bottlenecks, and the converse process of fusing
of surfaces constitute crucial ingredients for achieving percolating
configurations. In these $D=4$ ensembles, one can then model Yang-Mills
topological properties in addition to the confinement properties.
A sobering lesson of the present exploratory study is the proliferation
of modeling parameters in the type of formulation investigated here,
compared to previous models utilizing a hypercubic scaffolding. This
of course restricts the predictive potential of such models. Nevertheless,
it has proven possible to reproduce the qualitative features of
confinement physics seen in $SU(2)$ Yang-Mills theory within the
formulation constructed here.

\acknowledgments{We thank S.~Catterall and D.~A.~Johnston for helpful
discussions. This research was supported by the U.S. DOE through grant
DE-FG02-96ER40965 (D.A.,M.E.) and the Erwin Schr\"odinger Fellowship
program of the Austrian Science Fund FWF (``Fonds zur F\"orderung der
wissenschaftlichen Forschung'') under Contract No. J3425-N27 (R.H.).}

\bibliographystyle{utphys}
\bibliography{../literatur}

\providecommand{\href}[2]{#2}\begingroup\raggedright\begin{thebibliography}{10}

\bibitem{'tHooft:1977hy}
G.~'t~Hooft, ``On the phase transition towards permanent quark confinement,''
\href{http://dx.doi.org/10.1016/0550-3213(78)90153-0}{{\em Nucl. Phys.}
  {\bfseries B138} (1978) 1}.

\bibitem{Vinciarelli:1978kp}
P.~Vinciarelli, ``Fluxon solutions in nonabelian gauge models,''
\href{http://dx.doi.org/10.1016/0370-2693(78)90493-8}{{\em Phys. Lett.}
  {\bfseries B78} (1978) 485--488}.

\bibitem{Yoneya:1978dt}
T.~Yoneya, ``Z(n) topological excitations in yang-mills theories: Duality and
  confinement,''
\href{http://dx.doi.org/10.1016/0550-3213(78)90502-3}{{\em Nucl. Phys.}
  {\bfseries B144} (1978) 195}.

\bibitem{Cornwall:1979hz}
J.~M. Cornwall, ``{Quark confinement and Vortices in massive gauge invariant
  QCD},''
\href{http://dx.doi.org/10.1016/0550-3213(79)90111-1}{{\em Nucl. Phys.}
  {\bfseries B157} (1979) 392}.

\bibitem{Mack:1978rq}
G.~Mack and V.~B. Petkova, ``{Comparison of Lattice Gauge Theories with gauge
  groups Z(2) and SU(2)},''
\href{http://dx.doi.org/10.1016/0003-4916(79)90346-4}{{\em Ann. Phys.}
  {\bfseries 123} (1979) 442}.

\bibitem{Nielsen:1979xu}
H.~B. Nielsen and P.~Olesen, ``{A Quantum Liquid Model for the QCD Vacuum:
  Gauge and Rotational Invariance of Domained and Quantized Homogeneous Color
  Fields},''
\href{http://dx.doi.org/10.1016/0550-3213(79)90065-8}{{\em Nucl. Phys.}
  {\bfseries B160} (1979) 380}.

\bibitem{DelDebbio:1996mh}
{L. Del Debbio, M. Faber, J. Greensite, and {\v S}. Olejn\'{\i}k}, ``{Center
  dominance and Z(2) vortices in SU(2) lattice gauge theory},''
  \href{http://dx.doi.org/10.1103/PhysRevD.55.2298}{{\em Phys. Rev. D}
  {\bfseries 55} (1997) 2298--2306},
\href{http://arxiv.org/abs/9610005}{{\ttfamily arXiv:9610005 [hep-lat]}}.

\bibitem{Langfeld:1997jx}
K.~Langfeld, H.~Reinhardt, and O.~Tennert, ``{Confinement and scaling of the
  vortex vacuum of SU(2) lattice gauge theory},'' {\em Phys. Lett.} {\bfseries
  B419} (1998) 317--321,
\href{http://arxiv.org/abs/9710068}{{\ttfamily arXiv:9710068 [hep-lat]}}.

\bibitem{DelDebbio:1997ke}
{L. Del Debbio, and M. Faber, and J. Greensite, and {\v S}. Olejn\'{\i}k},
  ``Center dominance, center vortices, and confinement,''
\href{http://arxiv.org/abs/9708023}{{\ttfamily arXiv:9708023 [hep-lat]}}.

\bibitem{Langfeld:1998cz}
K.~Langfeld, O.~Tennert, M.~Engelhardt, and H.~Reinhardt, ``{Center vortices of
  Yang-Mills theory at finite temperatures},'' {\em Phys. Lett.} {\bfseries
  B452} (1999) 301,
\href{http://arxiv.org/abs/9805002}{{\ttfamily arXiv:9805002 [hep-lat]}}.

\bibitem{Kovacs:1998xm}
T.~G. Kovacs and E.~T. Tomboulis, ``Vortices and confinement at weak
  coupling,'' \href{http://dx.doi.org/10.1103/PhysRevD.57.4054}{{\em Phys. Rev.
  D} {\bfseries 57} (1998) 4054--4062},
\href{http://arxiv.org/abs/9711009}{{\ttfamily arXiv:9711009 [hep-lat]}}.

\bibitem{Engelhardt:1999wr}
M.~Engelhardt and H.~Reinhardt, ``{Center vortex model for the infrared sector
  of Yang-Mills theory: Confinement and deconfinement},''
  \href{http://dx.doi.org/10.1016/S0550-3213(00)00445-4}{{\em Nucl.Phys.}
  {\bfseries B585} (2000) 591--613},
\href{http://arxiv.org/abs/hep-lat/9912003}{{\ttfamily arXiv:hep-lat/9912003
  [hep-lat]}}.

\bibitem{Engelhardt:1999fd}
M.~Engelhardt, K.~Langfeld, H.~Reinhardt, and O.~Tennert, ``{Deconfinement in
  SU(2) Yang-Mills theory as a center vortex percolation transition},''
  \href{http://dx.doi.org/10.1103/PhysRevD.61.054504}{{\em Phys.Rev.}
  {\bfseries D61} (2000) 054504},
\href{http://arxiv.org/abs/hep-lat/9904004}{{\ttfamily arXiv:hep-lat/9904004
  [hep-lat]}}.

\bibitem{Bertle:2002mm}
R.~Bertle and M.~Faber, ``{Vortices, confinement and Higgs fields},''
\href{http://arXiv.org/abs/0212027}{{\ttfamily arXiv:0212027 [hep-lat]}}.

\bibitem{Engelhardt:2003wm}
M.~Engelhardt, M.~Quandt, and H.~Reinhardt, ``{Center vortex model for the
  infrared sector of SU(3) Yang-Mills theory: Confinement and deconfinement},''
  \href{http://dx.doi.org/10.1016/j.nuclphysb.2004.02.036}{{\em Nucl.Phys.}
  {\bfseries B685} (2004) 227--248},
\href{http://arxiv.org/abs/0311029}{{\ttfamily arXiv:0311029 [hep-lat]}}.

\bibitem{Hollwieser:2014lxa}
{R. H\"ollwieser, D. Altarawneh, and M. Engelhardt}, ``{Random center vortex
  lines in continuous 3D space-time},''
  \href{http://dx.doi.org/http://dx.doi.org/10.1063/1.4938613}{{\em AIP Conf.
  Proc.} {\bfseries 1701, ConfinementXI} (2014) 030007}.
  \url{http://scitation.aip.org/content/aip/proceeding/aipcp/10.1063/1.4938613}.

\bibitem{Altarawneh:2015bya}
{D. Altarawneh, R. H\"ollwieser and M. Engelhardt}, ``{Confining Bond
  Rearrangement in Random Center Vortex Models},'' {\em (accepted by PRD)}
  (2015) ,
\href{http://arxiv.org/abs/1508.07596}{{\ttfamily arXiv:1508.07596 [hep-lat]}}.

\bibitem{Hollwieser:2015qea}
{R. H\"ollwieser and D. Altarawneh}, ``{Center Vortices, Area Law and the
  Catenary Solution},'' \href{http://dx.doi.org/10.1142/S0217751X15502073}{{\em
  Int. J. Mod. Phys.} {\bfseries A30} (2015) 1550207},
\href{http://arxiv.org/abs/1509.00145}{{\ttfamily arXiv:1509.00145 [hep-lat]}}.

\bibitem{Greensite:2003bk}
J.~Greensite, ``The confinement problem in lattice gauge theory,''
  \href{http://dx.doi.org/10.1016/S0146-6410(03)90012-3}{{\em Prog. Part. Nucl.
  Phys.} {\bfseries 51} (2003) 1},
\href{http://arxiv.org/abs/hep-lat/0301023}{{\ttfamily arXiv:hep-lat/0301023
  [hep-lat]}}.

\bibitem{Greensite:2014gra}
{J. Greensite and R. H\"ollwieser}, ``{Double-winding Wilson loops and monopole
  confinement mechanisms},''
  \href{http://dx.doi.org/10.1103/PhysRevD.91.054509}{{\em Phys. Rev.}
  {\bfseries D91} no.~5, (2015) 054509},
\href{http://arxiv.org/abs/1411.5091}{{\ttfamily arXiv:1411.5091 [hep-lat]}}.

\bibitem{Bertle:2001xd}
R.~Bertle, M.~Engelhardt, and M.~Faber, ``{Topological susceptibility of
  Yang-Mills center projection vortices},'' {\em Phys. Rev. D} {\bfseries 64}
  (2001) 074504,
\href{http://arxiv.org/abs/0104004}{{\ttfamily arXiv:0104004 [hep-lat]}}.

\bibitem{Engelhardt:2000wc}
M.~Engelhardt, ``{Center vortex model for the infrared sector of Yang-Mills
  theory: Topological susceptibility},''
  \href{http://dx.doi.org/10.1016/S0550-3213(00)00350-3}{{\em Nucl.Phys.}
  {\bfseries B585} (2000) 614},
\href{http://arxiv.org/abs/0004013}{{\ttfamily arXiv:0004013 [hep-lat]}}.

\bibitem{Engelhardt:2010ft}
M.~Engelhardt, ``{Center vortex model for the infrared sector of SU(3)
  Yang-Mills theory: Topological susceptibility},''
  \href{http://dx.doi.org/10.1103/PhysRevD.83.025015}{{\em Phys. Rev. D}
  {\bfseries 83} (2011) 025015},
\href{http://arxiv.org/abs/1008.4953}{{\ttfamily arXiv:1008.4953 [hep-lat]}}.

\bibitem{Hollwieser:2010mj}
{R. H\"ollwieser, M. Faber, U.M. Heller}, ``{Lattice Index Theorem and
  Fractional Topological Charge},''
\href{http://arxiv.org/abs/1005.1015}{{\ttfamily arXiv:1005.1015 [hep-lat]}}.

\bibitem{Hollwieser:2011uj}
{R. H\"ollwieser, M. Faber, U.M. Heller}, ``{Intersections of thick Center
  Vortices, Dirac Eigenmodes and Fractional Topological Charge in SU(2) Lattice
  Gauge Theory},'' \href{http://dx.doi.org/10.1007/JHEP06(2011)052}{{\em JHEP}
  {\bfseries 1106} (2011) 052},
\href{http://arxiv.org/abs/1103.2669}{{\ttfamily arXiv:1103.2669 [hep-lat]}}.

\bibitem{Schweigler:2012ae}
{T. Schweigler, R. H\"ollwieser, M. Faber and U.M. Heller}, ``{Colorful SU(2)
  center vortices in the continuum and on the lattice},''
  \href{http://dx.doi.org/10.1103/PhysRevD.87.054504}{{\em Phys.Rev.}
  {\bfseries D87} no.~5, (2013) 054504},
\href{http://arxiv.org/abs/1212.3737}{{\ttfamily arXiv:1212.3737 [hep-lat]}}.

\bibitem{Hollwieser:2012kb}
{R. H\"ollwieser, M. Faber, U.M. Heller}, ``{Critical analysis of topological
  charge determination in the background of center vortices in SU(2) lattice
  gauge theory},'' \href{http://dx.doi.org/10.1103/PhysRevD.86.014513}{{\em
  Phys. Rev. D} {\bfseries 86} (2012) 014513},
\href{http://arxiv.org/abs/1202.0929}{{\ttfamily arXiv:1202.0929 [hep-lat]}}.

\bibitem{Hollwieser:2014mxa}
{R. H\"ollwieser and M. Engelhardt}, ``{Smearing Center Vortices},'' {\em PoS}
  {\bfseries LAT2014} (2014) 356,
\href{http://arxiv.org/abs/1411.7097}{{\ttfamily arXiv:1411.7097 [hep-lat]}}.

\bibitem{Hollwieser:2015koa}
{R. H\"ollwieser and M. Engelhardt}, ``{Approaching SU(2) gauge dynamics with
  smeared Z(2) vortices},'' {\em Phys. Rev.} {\bfseries D92} (2015) 034502,
\href{http://arxiv.org/abs/1503.00016}{{\ttfamily arXiv:1503.00016 [hep-lat]}}.

\bibitem{deForcrand:1999ms}
P.~de~Forcrand and M.~D'Elia, ``{On the relevance of center vortices to QCD},''
  \href{http://dx.doi.org/10.1103/PhysRevLett.82.4582}{{\em Phys. Rev. Lett.}
  {\bfseries 82} (1999) 4582--4585},
\href{http://arxiv.org/abs/hep-lat/9901020}{{\ttfamily arXiv:hep-lat/9901020
  [hep-lat]}}.

\bibitem{Alexandrou:1999vx}
C.~Alexandrou, P.~de~Forcrand, and M.~D'Elia, ``{The role of center vortices in
  QCD},'' \href{http://dx.doi.org/10.1016/S0375-9474(99)00763-0}{{\em Nucl.
  Phys.} {\bfseries A663} (2000) 1031--1034},
\href{http://arxiv.org/abs/hep-lat/9909005}{{\ttfamily arXiv:hep-lat/9909005
  [hep-lat]}}.

\bibitem{Engelhardt:1999xw}
M.~Engelhardt and H.~Reinhardt, ``{Center projection vortices in continuum
  Yang-Mills theory},'' {\em Nucl. Phys.} {\bfseries B567} (2000) 249,
\href{http://arxiv.org/abs/9907139}{{\ttfamily arXiv:9907139 [hep-th]}}.

\bibitem{Reinhardt:2000ck2}
H.~Reinhardt and M.~Engelhardt, ``Center vortices in continuum yang-mills
  theory,'' in {\em Quark Confinement and the Hadron Spectrum IV}, W.~Lucha and
  K.~M. Maung, eds., pp.~150--162.
\newblock World Scientific, 2002.
\newblock
\href{http://arxiv.org/abs/0010031}{{\ttfamily arXiv:0010031 [hep-th]}}.
\newblock

\bibitem{Engelhardt:2002qs}
M.~Engelhardt, ``{Center vortex model for the infrared sector of Yang-Mills
  theory: Quenched Dirac spectrum and chiral condensate},''
  \href{http://dx.doi.org/10.1016/S0550-3213(02)00470-4}{{\em Nucl.Phys.}
  {\bfseries B638} (2002) 81--110},
\href{http://arxiv.org/abs/hep-lat/0204002}{{\ttfamily arXiv:hep-lat/0204002
  [hep-lat]}}.

\bibitem{Leinweber:2006zq}
D.~Leinweber, P.~Bowman, U.~Heller, D.~Kusterer, K.~Langfeld, {\em et~al.},
  ``{Role of centre vortices in dynamical mass generation},''
\href{http://dx.doi.org/10.1016/j.nuclphysbps.2006.08.065}{{\em
  Nucl.Phys.Proc.Suppl.} {\bfseries 161} (2006) 130--135}.

\bibitem{Bornyakov:2007fz}
{V. Bornyakov {\it et al.}}, ``{Interrelation between monopoles, vortices,
  topological charge and chiral symmetry breaking: Analysis using overlap
  fermions for SU(2)},''
  \href{http://dx.doi.org/10.1103/PhysRevD.77.074507}{{\em Phys. Rev. D}
  {\bfseries 77} (2008) 074507},
\href{http://arxiv.org/abs/0708.3335}{{\ttfamily arXiv:0708.3335 [hep-lat]}}.

\bibitem{Hollwieser:2008tq}
{R. H\"ollwieser, M. Faber, J. Greensite, U.M. Heller, and {\v S}.
  Olejn\'{\i}k}, ``{Center Vortices and the Dirac Spectrum},''
  \href{http://dx.doi.org/10.1103/PhysRevD.78.054508}{{\em Phys. Rev. D}
  {\bfseries 78} (2008) 054508},
\href{http://arxiv.org/abs/0805.1846}{{\ttfamily arXiv:0805.1846 [hep-lat]}}.

\bibitem{Bowman:2010zr}
P.~O. Bowman, K.~Langfeld, D.~B. Leinweber, A.~Sternbeck, L.~von Smekal, {\em
  et~al.}, ``{Role of center vortices in chiral symmetry breaking in SU(3)
  gauge theory},'' \href{http://dx.doi.org/10.1103/PhysRevD.84.034501}{{\em
  Phys.Rev.} {\bfseries D84} (2011) 034501},
\href{http://arxiv.org/abs/1010.4624}{{\ttfamily arXiv:1010.4624 [hep-lat]}}.

\bibitem{Hollwieser:2013xja}
{R. H\"ollwieser, T. Schweigler, M. Faber and U.M. Heller}, ``{Center Vortices
  and Chiral Symmetry Breaking in SU(2) Lattice Gauge Theory},''
  \href{http://dx.doi.org/10.1103/PhysRevD.88.114505}{{\em Phys.Rev.}
  {\bfseries D88} (2013) 114505},
\href{http://arxiv.org/abs/1304.1277}{{\ttfamily arXiv:1304.1277 [hep-lat]}}.

\bibitem{Brambilla:2014jmp}
N.~Brambilla, S.~Eidelman, P.~Foka, S.~Gardner, A.~Kronfeld, {\em et~al.},
  ``{QCD and Strongly Coupled Gauge Theories: Challenges and Perspectives},''
  {\em EJPC} {\bfseries 74} (2014) {Issue 10},
\href{http://arxiv.org/abs/1404.3723}{{\ttfamily arXiv:1404.3723 [hep-ph]}}.

\bibitem{Hollwieser:2014osa}
{R. H\"ollwieser, M. Faber, Th. Schweigler, and U.M. Heller}, ``{Chiral
  Symmetry Breaking from Center Vortices},'' {\em PoS} {\bfseries LAT2013}
  (2014) 505,
\href{http://arxiv.org/abs/1410.2333}{{\ttfamily arXiv:1410.2333 [hep-lat]}}.

\bibitem{Trewartha:2014ona}
D.~Trewartha, W.~Kamleh, and D.~Leinweber, ``{Centre Vortex Effects on the
  Overlap Quark Propagator},'' {\em PoS} {\bfseries LATTICE2014} (2014) 357,
\href{http://arxiv.org/abs/1411.0766}{{\ttfamily arXiv:1411.0766 [hep-lat]}}.

\bibitem{Trewartha:2015nna}
D.~Trewartha, W.~Kamleh, and D.~Leinweber, ``{Evidence that centre vortices
  underpin dynamical chiral symmetry breaking in SU(3) gauge theory},''
  \href{http://dx.doi.org/10.1016/j.physletb.2015.06.025}{{\em Phys. Lett.}
  {\bfseries B747} (2015) 373--377},
\href{http://arxiv.org/abs/1502.06753}{{\ttfamily arXiv:1502.06753 [hep-lat]}}.

\bibitem{Quandt:2004gy}
M.~Quandt, H.~Reinhardt, and M.~Engelhardt, ``{Center vortex model for the
  infrared sector of SU(3) Yang-Mills theory - vortex free energy},''
  \href{http://dx.doi.org/10.1103/PhysRevD.71.054026}{{\em Phys.Rev.}
  {\bfseries D71} (2005) 054026},
\href{http://arxiv.org/abs/hep-lat/0412033}{{\ttfamily arXiv:hep-lat/0412033
  [hep-lat]}}.

\bibitem{Engelhardt:2004qq}
M.~Engelhardt, ``{Center vortex model for the infrared sector of SU(3)
  Yang-Mills theory - baryonic potential},''
  \href{http://dx.doi.org/10.1103/PhysRevD.70.074004}{{\em Phys. Rev. D}
  {\bfseries 70} (2004) 074004},
\href{http://arxiv.org/abs/0406022}{{\ttfamily arXiv:0406022 [hep-lat]}}.

\bibitem{Engelhardt:2005qu}
M.~Engelhardt, ``{Center vortex model for the infrared sector of SU(4)
  Yang-Mills theory: String tensions and deconfinement transition},''
  \href{http://dx.doi.org/10.1103/PhysRevD.73.034015}{{\em Phys. Rev. D}
  {\bfseries 73} (2006) 034015},
\href{http://arxiv.org/abs/0512015}{{\ttfamily arXiv:0512015 [hep-lat]}}.

\bibitem{Engelhardt:2006ep}
M.~Engelhardt and B.~Sperisen, ``{Center vortex model for Sp(2) Yang-Mills
  theory},'' \href{http://dx.doi.org/10.1103/PhysRevD.74.125011}{{\em
  Phys.Rev.} {\bfseries D74} (2006) 125011},
\href{http://arxiv.org/abs/hep-lat/0610074}{{\ttfamily arXiv:hep-lat/0610074
  [hep-lat]}}.

\bibitem{Baillie:1989mv}
C.~Baillie, D.~Johnston, and R.~Williams, ``{Computational aspects of
  simulating dynamically triangulated random surfaces},''
  \href{http://dx.doi.org/http://dx.doi.org/10.1016/0010-4655(90)90139-R}{{\em
  Comp. Phys. Commun.} {\bfseries 58} no.~1, (1990) 105}.
  \url{http://www.sciencedirect.com/science/article/pii/001046559090139R}.

\bibitem{Catterall:1995:SDT}
S.~Catterall, ``Simulations of dynamically triangulated gravity -- an algorithm
  for arbitrary dimension,''
  \href{http://dx.doi.org/http://dx.doi.org/10.1016/0010-4655(94)00117-K}{{\em
  Comp. Phys. Commun.} {\bfseries 87} no.~3, (June, 1995) 409--415}.
  \url{http://www.sciencedirect.com/science/article/pii/001046559400117K}.

\bibitem{Jain:1992bs}
S.~Jain and S.~D. Mathur, ``{World-sheet geometry and baby universes in 2-D
  quantum gravity},''
  \href{http://dx.doi.org/10.1016/0370-2693(92)91769-6}{{\em Phys. Lett.}
  {\bfseries B286} (1992) 239--246},
\href{http://arxiv.org/abs/hep-th/9204017}{{\ttfamily arXiv:hep-th/9204017
  [hep-th]}}.

\bibitem{Ambjorn:1993vz}
J.~Ambj{\o}rn, S.~Jain, and G.~Thorleifsson, ``{Baby universes in 2-d quantum
  gravity},'' \href{http://dx.doi.org/10.1016/0370-2693(93)90188-N}{{\em Phys.
  Lett.} {\bfseries B307} (1993) 34--39},
\href{http://arxiv.org/abs/hep-th/9303149}{{\ttfamily arXiv:hep-th/9303149
  [hep-th]}}.

\bibitem{Thorleifsson:1995ki}
G.~Thorleifsson and S.~Catterall, ``{A real space renormalization group for
  random surfaces},''
  \href{http://dx.doi.org/10.1016/0550-3213(95)00664-8}{{\em Nucl. Phys.}
  {\bfseries B461} (1996) 350--370},
\href{http://arxiv.org/abs/hep-lat/9510003}{{\ttfamily arXiv:hep-lat/9510003
  [hep-lat]}}.

\bibitem{Ambjorn:1995dj}
J.~Ambj{\o}rn and J.~Jurkiewicz, ``{Scaling in four-dimensional quantum
  gravity},'' \href{http://dx.doi.org/10.1016/0550-3213(95)00303-A}{{\em Nucl.
  Phys.} {\bfseries B451} (1995) 643--676},
\href{http://arxiv.org/abs/hep-th/9503006}{{\ttfamily arXiv:hep-th/9503006
  [hep-th]}}.

\end{thebibliography}\endgroup

\end{document}